\documentclass[12pt]{article}
\usepackage{epsfig}

\newcommand\pubnumber{SLAC--PUB--10440}
\newcommand\pubdate{May, 2004}
\newcommand\hepnumber{hep-ph/0405214}

\def\SISSA{Scuola Internazionale di Studi Avanzati (SISSA) \\ 
 via Beirut 4, 
 I-34014, Trieste, ITALY}
\def\SLAC{Stanford Linear Accelerator Center\\
    2575 Sand Hill Road, Menlo Park, California 94025 USA}
\def\doeack{\footnote{Work supported by the Department of Energy,
                     contract DE--AC03--76SF00515.}}

\def\Title#1{\begin{center} {\Large #1 } \end{center}}
\def\Author#1{\begin{center}{ \sc #1} \end{center}}
\def\Address#1{\begin{center}{ \it #1} \end{center}}

\def\submit#1{\begin{center}Submitted to {\sl #1} \end{center}}
\newcommand\pubblock{\rightline{\begin{tabular}{l} \pubnumber\\
         \pubdate \\ \hepnumber \end{tabular}}}
\newenvironment{Abstract}{\begin{quotation} \begin{center}
                       ABSTRACT
     \end{center}\bigskip  }{\end{quotation}}

\def\submit#1{\begin{center}Submitted to {\sl #1} \end{center}}
\def\Acknowledgements{\bigskip  \bigskip \begin{center} \begin{large}
             \bf ACKNOWLEDGEMENTS \end{large}\end{center}}

\def\beq{\begin{equation}}
\def\eeq#1{\label{#1}\end{equation}}
\def\eeqn{\end{equation}}

\newenvironment{Eqnarray}%
   {\arraycolsep 0.14em\begin{eqnarray}}{\end{eqnarray}}
\def\beqa{\begin{Eqnarray}}
\def\eeqa#1{\label{#1}\end{Eqnarray}}
\def\eeqan{\end{Eqnarray}}
\def\CR{\nonumber \\ }

\def\leqn#1{(\ref{#1})}

\let\bar=\overbar

\def\VEV#1{\left\langle{ #1} \right\rangle}

\def\lsim{\mathrel{\raise.3ex\hbox{$<$\kern-.75em\lower1ex\hbox{$\sim$}}}}
\def\gsim{\mathrel{\raise.3ex\hbox{$>$\kern-.75em\lower1ex\hbox{$\sim$}}}}

\def\Im{{\rm Im}}

\def\L{{\cal L}}

\def\half{\frac{1}{2}}

\def\del{\partial}
\def\Dslash{\not{\hbox{\kern-4pt $D$}}}
\def\dslash{\not{\hbox{\kern-2pt $\del$}}}

\def\eff{{\mbox{\scriptsize eff}}}

\def\GUT{{\mbox{\scriptsize GUT}}}

\def\msb{{\bar{\scriptsize M \kern -1pt S}}}

\def\drb{{\bar{\scriptsize D \kern -1pt R}}}

\def\s#1{\widetilde{#1}}

\makeatletter
\def\section{\@startsection{section}{0}{\z@}{5.5ex plus .5ex minus
 1.5ex}{2.3ex plus .2ex}{\large\bf}}
\def\subsection{\@startsection{subsection}{1}{\z@}{3.5ex plus .5ex minus
 1.5ex}{1.3ex plus .2ex}{\normalsize\bf}}
\def\subsubsection{\@startsection{subsubsection}{2}{\z@}{-3.5ex plus
-1ex minus  -.2ex}{2.3ex plus .2ex}{\normalsize\sl}}

\renewcommand{\@makecaption}[2]{%
   \vskip 10pt
   \setbox\@tempboxa\hbox{\small #1: #2}
   \ifdim \wd\@tempboxa >\hsize     
       \small #1: #2\par          
     \else                        
       \hbox to\hsize{\hfil\box\@tempboxa\hfil}
   \fi}

 \def\citenum#1{{\def\@cite##1##2{##1}\cite{#1}}}
 
\newcount\@tempcntc
\def\@citex[#1]#2{\if@filesw\immediate\write\@auxout{\string\citation{#2}}\fi
  \@tempcnta\z@\@tempcntb\m@ne\def\@citea{}\@cite{\@for\@citeb:=#2\do
    {\@ifundefined
       {b@\@citeb}{\@citeo\@tempcntb\m@ne\@citea\def\@citea{,}{\bf ?}\@warning
       {Citation `\@citeb' on page \thepage \space undefined}}%
    {\setbox\z@\hbox{\global\@tempcntc0\csname b@\@citeb\endcsname\relax}%
     \ifnum\@tempcntc=\z@ \@citeo\@tempcntb\m@ne
       \@citea\def\@citea{,}\hbox{\csname b@\@citeb\endcsname}%
     \else
      \advance\@tempcntb\@ne
      \ifnum\@tempcntb=\@tempcntc
      \else\advance\@tempcntb\m@ne\@citeo
      \@tempcnta\@tempcntc\@tempcntb\@tempcntc\fi\fi}}\@citeo}{#1}}
\def\@citeo{\ifnum\@tempcnta>\@tempcntb\else\@citea\def\@citea{,}%
  \ifnum\@tempcnta=\@tempcntb\the\@tempcnta\else
  {\advance\@tempcnta\@ne\ifnum\@tempcnta=\@tempcntb \else\def\@citea{--}\fi
    \advance\@tempcnta\m@ne\the\@tempcnta\@citea\the\@tempcntb}\fi\fi}
\makeatother

\def\GUT{{\mbox{\scriptsize GUT}}}
\def\MGUT{M_\GUT}
\def\SUSY{{\mbox{\scriptsize SUSY}}}
\def\MSUSY{M_\SUSY}
\def\L{{\cal L}}

\begin{document}
\begin{titlepage}
\pubblock
\setcounter{footnote}{2}

\vfill
\Title{The Contribution from Neutrino Yukawa Couplings to
Lepton Electric Dipole Moments}
\vfill
\Author{Yasaman Farzan$^{1,2,3}$ and Michael E. Peskin$^{1,}$\doeack}
\Address{$^1$\SLAC \\ $^2$\SISSA}
\vfill
\begin{Abstract}
To explain the observed neutrino masses through the seesaw mechanism, a
supersymmetric generalization of the Standard Model should  include 
heavy right-handed neutrino supermultiplets.  
Then the neutrino Yukawa couplings
can induce CP violation in the lepton sector.  In this paper, we 
compute the contribution of these CP violating terms to 
lepton electric dipole moments.  We introduce a 
new formalism that makes use of supersymmetry to expose the GIM cancellations.
In the region of small $\tan\beta$, we find a different result from that
given previously by Ellis, Hisano, Raidal, and Shimizu.  We confirm the 
structure found by this group, but with a much smaller overall coefficient.
In the region of large $\tan\beta$, we recompute the leading term that has
been identified by Masina and confirm her result.
We discuss the implications of these results 
for constraints on the $Y_\nu$.
\end{Abstract}
\vfill
\vfill
\submit{Physical Review {\bf D}}
\vfill
\vfill
\end{titlepage}
\def\thefootnote{\fnsymbol{footnote}}
\setcounter{footnote}{0}
\tableofcontents
\newpage

\section{Introduction}

The discovery of neutrino mass has not only required revision of the 
Standard Model of particle physics but also of theories that go beyond
the Standard Model.  A compelling idea  for the origin of the observed small 
neutrino masses is the seesaw mechanism.  This requires the introduction 
of heavy singlet leptons, that is, right-handed neutrinos.  In the 
context of supersymmetric 
theories, these singlet leptons belong to new chiral supermultiplets $N_i$,
one for each fermion generation.  The Yukawa couplings and soft supersymmetry
breaking terms associated with these right-handed neutrino supermultiplets 
can play important roles in lepton flavor violating 
processes~\cite{antonio} and in the production of the
baryon number of the universe through 
leptogenesis~\cite{yanagida}.

A particularly important aspect of this model is the appearance of new
sources of CP violation.  In addition to new CP violating parameters generic
to new physics---in supersymmetry, for example, the phases of $\mu$ and 
the $A$---new phases are possible in 
the neutrino Yukawa couplings and in the neutrino $B$ term ($B M 
\tilde{N} \tilde{N}$).
Complex Yukawa couplings can lead to observable CP violation in 
neutrino oscillations, and all of these parameters can be the source of the 
CP 
violation 
that generated a fermion-antifermion asymmetry in the early 
universe~\cite{yanagida,D'A}.

To test whether the observed matter-antimatter asymmetry indeed arose 
from leptogenesis, it is necessary to determine the CP violating phases 
from microscopic measurements.  There has been much analysis of 
CP violating observables in neutrino mixing.   
In principle, it  is possible to determine 
all seesaw parameters studying neutrino and sneutrino 
mass matrices but, in practice, it will be quite 
challenging, if possible at all, to extract all of 
these parameters in the near future~\cite{hopeless}.  
Another possible experimental
approach to test CP violation in the lepton sector is to measure 
 the electric dipole moments (EDMs) of 
charged leptons~\cite{dutta}.  There are  in fact
many 
possible ways that underlying CP violating couplings could give rise to 
lepton EDMs.  Thus, it is important to classify these effects
and, if possible, to learn how to separate them from one another.

If CP violation is provided by phases of soft supersymmetry breaking 
parameters, it is straightforward to generate a contribution to 
lepton electric dipole moments in one-loop order.   This possibility
has been explored by many authors~\cite{amu}. However, it is also possible
to generate lepton EDMs in models in which the soft supersymmetry breaking
terms are CP conserving, by making use of phases in the neutrino 
Yukawa couplings.  A particularly simple context to study this effect is
to consider models in which the soft supersymmetry breaking scalar masses
are exactly flavor-universal and the $A$ terms are exactly proportional
to the Yukawa couplings.  Such models arise in the simplest paradigms for
gravity-mediated supersymmetry breaking~\cite{AN}.   The idea of `gaugino
mediation' provides an attractive 
way to realize this scheme in the context of a complete
unified or superstring model~\cite{gauginomed}.

In this class of models with universal soft supersymmetry breaking 
interactions, the flavor and CP violation due to 
the neutrino Yukawa couplings is computed by 
integrating out the right-handed neutrino $N_i$ superfields.  This particular
model has been studied in a number of papers, beginning with \cite{HNY}.
In particular, the contribution of neutrino Yukawa couplings to lepton EDMs
has been studied in this context by Romanino and Strumia~\cite{RandS},
Ellis, Hisano, Raidal, and 
Shimizu (EHRS)~\cite{ellisteam}, and  
Masina~\cite{isabella}. These authors found that the analysis is complicated
by GIM cancellations, so that the first nonzero contribution to the EDMs 
arises in two-loop order and has the form of a commutator of different
combinations of the Yukawa matrices.

The work of \cite{ellisteam} and \cite{isabella} used a 
renormalization group equation (RGE)
approach to evaluate the leading logarithmic contributions to the lepton EDMs.
We thought that it might be valuable to extend these calculations by 
evaluating the complete contribution 
to the lepton EDMs without making leading-log approximations.  In this paper,
we present a new accounting method for the CP violating effects of the 
right-handed neutrino sector that makes this calculation straightforward.
Our results, however, differ from those of  \cite{ellisteam}
 even at the leading-log level.  We confirm the general 
structure of
the answers found by this group---in particular, the commutator structure
noted in the previous paragraph.  However, we claim that there are further
cancellations not found in their papers that one must resolve to obtain
the correct detailed formulae.  We confirm Masina's result for the
large $\tan\beta$ region, up to some minor factors, using a method that is
much more transparent.

The outline of this paper is then as follows:  In Section 2, we specify
the model in which we are working.  In Section 3, we describe our
procedure for integrating out the $N_i$ supermultiplets and identifying 
CP violating contributions.  In Section 4, we carry out this procedure
for the leading CP violating contribution proportional to $Y_\nu^4$, where
$Y_\nu$ is the neutrino Yukawa coupling.  We find a result that is 
parametrically smaller than that of EHRS by one power of a large logarithm.
In Section 5, we reconsider the analysis of EHRS and show how that logarithm
cancels out using their method.  In Section 6, we give a formula for
the lepton EDMs that arises from this contribution.

In \cite{isabella}, Masina
 pointed out that, for large values of $\tan\beta$, a 
different contribution can dominate the evaluation of the lepton EDMs.
This new term arises at one higher loop order, at order $Y_\nu^4Y_\ell^2$, 
where $Y_\ell$ is the charged lepton Yukawa coupling. 
In Section 7, we evaluate this contribution, which requires a nontrivial
two-loop diagram calculation.

In Section 8, we make numerical estimates of the electron EDM from our
new formulae and compare these to the results of other models of lepton 
CP violation.

Our calculations in this paper look specifically at the terms resulting 
from integrating out the right-handed neutrino sector.  We work from an
 initial assumption that the soft supersymmetry breaking terms are universal
and flavor-independent.  In a model with renormalizable interactions that 
violate the flavor and CP symmetries, this initial condition is not technically
natural.  Thus, there will in general
 be other CP violating contributions, for example,
from the thresholds at the grand unification scale $\MGUT$, 
that should be 
added to the formulae we present here.  Because, in all of our formulae,
the leading logarithmic behavior cancels
due to a GIM cancellation, our terms are not parametrically enhanced over
those from the GUT threshold.  In specific models, the GUT scale terms can
be numerically smaller than the terms from the right-handed neutrino scale;
the authors of \cite{HNY}, for example, argue this for their $SU(5)$ GUT
model.  In 
any event, our formulae are computed precisely for the effective theory 
of Section 2 with minimal subtraction (in the $\bar{DR}$ scheme) at $\MGUT$.
By noting this prescription,
it should be straightforward to add GUT threshold corrections to our
results when these are computed in a particular GUT model.

\section{The model}

We consider the supersymmetric Standard Model coupled to three chiral
supermultiplets $N_i$ which contain the heavy right-handed neutrinos 
associated
with the seesaw mechanism.  The superpotential of the model contains the 
following terms involving lepton supermultiplets:
\beq
   W =  Y^{ik}_\ell \epsilon_{\alpha\beta} 
H_{1\alpha} E_i L_{j\beta}
        -  Y^{ij}_\nu\epsilon_{\alpha\beta} H_{2\alpha} N_i L_{j\beta}
        - \mu \epsilon_{\alpha\beta} 
H_{1\alpha}H_{2\beta}+ \half M_{ij} N_i N_j \ .
\eeq{theW}
In this equation, $L_{j\beta}$ is the supermultiplet containing the 
left-handed lepton fields $(\nu_{jL}, \ell^-_{jL})_\beta$, $E_i$ is the
superfield whose left-handed fermion is $\ell^+_{iL}$, and $N_i$ is the 
superfield whose left-handed fermion is $\bar\nu_{iL}$.  The $N_i$
are singlets of $SU(2)\times U(1)$.  We introduce 
the right-handed neutrino masses $M_{ij}$
 by hand, and we do not assume any {\it a priori} relation 
of these parameters
to the other couplings in \leqn{theW}.

Without loss of generality, we
can choose the basis and phases of $L$, $E$, and $N$ 
such that $M_{ij}$
and $Y_\ell^{ij}$ are real and diagonal.  We will refer to the diagonal
elements of these matrices as $M_i$, $Y_{\ell i}$.  These choices 
exhaust the freedom to redefine fields, and so the matrix $Y_\nu^{ij}$ is 
in general off-diagonal and complex.  The mass matrix of 
light neutrinos is given by 
\beq
        (m_\nu)_{ij}  =  \sum_k { Y_\nu^{ki} Y_\nu^{kj} 
\over M_k} \VEV{H^0_2}^2\ .
\eeq{mnus}
If the neutrino Yukawa couplings $Y_\nu^{ik}$ are of order 1, the 
requirement of small neutrino masses 
($m_\nu \sim 0.1$ eV) leads to large values of 
the $M_k$, of the order of $10^{14}$ GeV.

To the Lagrangian generated by \leqn{theW}, we must add appropriate 
soft supersymmetry breaking interactions.  
In this paper, we 
will assume that 
slepton masses are universal at the messenger scale (of the order of 
$\MGUT$) and that 
$A$ terms are strictly proportional to the 
corresponding Yukawa couplings,
with a real constant of proportionality.  We will assume that the 
phases of $\mu$ and of the gaugino masses are zero.
If these conditions are not met, it is possible to generate EDMs from 
one-loop diagrams, a possibility that has been exhaustively explored in the
literature~\cite{amu}.

This restriction to universal, CP invariant, flavor invariant soft 
supersymmetry breaking terms is not a natural restriction of the model
in the technical sense. It is violated by loop corrections due to the
neutrino Yukawa couplings.  In fact, our analysis in this paper is 
 to calculate the  CP violation induced by these corrections.
Consequently, the effects we find can be cut-off dependent.  As we have 
explained in the introduction, we will impose the universality and flavor
symmetry of the soft supersymmetry breaking interactions as an initial
condition, defined by minimal subtraction in the $\bar{DR}$ scheme at
$ \MGUT$.

With this prescription, we will 
 take the soft supersymmetry breaking terms for the lepton
sector to be 
\beqa
 \L_{SSB} &=& - m_0^2 \sum_f \s{f}^*\s{f} -  m_a \bar \lambda_a \lambda_a  
  -  a_0 \left(  Y_{\ell i} \epsilon_{\alpha\beta} H_{1\alpha} \tilde{E}_i 
\tilde{L}_{i\beta}
        -  Y^{ij}_\nu\epsilon_{\alpha\beta} H_{2\alpha} \tilde{N}_i 
\tilde{L}_{j\beta}
         \right)\CR & &  \hskip 0.2in
 -   (\half B_\nu M_i (\tilde{N}_i)^2 + 
H.c.) -(\half b_H \mu H_1H_2+ H.c.)
\eeqa{softterms}
where $\tilde{f}$ collectively represents 
sfermions, and we assume that $a_0$, $b_H$ and 
$B_\nu$ are 
real parameters.  The parameters $m_0$, $m_a$, $a_0$, 
and $b_H$ all have 
the dimensions of mass and are of order $\MSUSY \sim 
100~{\rm GeV}-1$ TeV.
CP violating phases arise both from 
the neutrino Yukawa couplings and from the neutrino $A$ term, but, in this
model, they are controlled by the same parameters.   
We should note that 
if any of the parameters $a_0$, $b_H$ or $B_\nu$ has an 
imaginary 
part, the corresponding  term can give a large 
contribution to 
lepton EDMs.  This point is discussed in some detail 
elsewhere~\cite{amu}.  The specific effects of the $B_\nu$ 
term have been analyzed in~\cite{YasamanB}

In computing the effects of the $N_i$ supermultiplets, it is convenient to 
work in components, keeping the auxiliary fields (the $F$ 
fields)  as independent fields.
We use two-component notation for the fermion fields.
With the effects of the Majorana mass term included, the propagators for the
 component fields of the $N_i$ take the form  
\beqa
 \VEV{ \tilde{N}_j(q) \tilde{N}_k^*(-q)}  =   { i \over q^2 -
M_j^2}\delta_{jk} & \qquad &
\VEV{ \tilde{N}_j(q) F_{N_k}(-q)}  =   { -i M_j \over q^2 -
M_j^2}\delta_{jk} \CR
\VEV{ N_{j}(q) N^\dagger_{k}(-q)}  =
            { i \sigma \cdot q \over q^2 - M_j^2}\delta_{jk}  & \qquad&
\VEV{ N_{j}(q) (N^{k}(-q))^T}  =
             {- i M_j c \over q^2 - M_j^2} \delta_{jk} \CR
  \VEV{ F_{N_j}(q) F_{N_k}^*(-q)}  &=&   { i q^2 \over q^2 - M_j^2}
\delta_{jk}   
\eeqa{theprops}
where $\sigma^\mu = (1,\vec \sigma)^\mu$ and $c = -i\sigma^2$ are $2\times 2$
components of the Dirac matrices and the charge conjugation matrix.

\section{Radiative corrections due to $Y_\nu$}

As it is well known, radiative corrections will distort
the form of Eq. \leqn{softterms}  and break the exact mass 
degeneracy between the sfermions.
In this section, we will focus on those radiative corrections 
to the parameters of Eq.  \leqn{softterms} that can induce CP-violating 
phase and EDMs, in particular, the 
effects of 
diagrams involving the neutrino Yukawa and $A$ terms.  We will discuss
the form of the effective Lagrangian  at scales just below the right-handed
neutrino mass scale.  
When we compute the induced EDMs in Section 6 and 7,
we will need to take into account some additional effects that come from
renormalization group running down to the electroweak scale.
In our analysis, we will always assume that
the right-handed neutrino masses $M_k$ are much larger than the supersymmetry
breaking mass terms, of order $\MSUSY$, so that any contribution 
suppressed by 
$\MSUSY/M_k$ can be neglected.   In this limit, the calculation that integrates
out the right-handed neutrino sector divides neatly into
a part that corrects the
supersymmetric Lagrangian and a part that corrects
the supersymmetry breaking perturbations.

First, we consider the radiative corrections to the 
supersymmetric part of the  Lagrangian.   We begin by noting that, to 
a good approximation, 
we can neglect diagrams that include vertices from the supersymmetry breaking
terms.  Except for the $\mu$ term, all coefficients in the supersymmetric
Lagrangian are dimensionless, while all supersymmetry breaking terms 
have coefficients with mass parameters of order 1 TeV or smaller.  
Therefore, corrections 
to the dimensionless coefficients 
from the supersymmetry breaking terms are at most of order of $\MSUSY/M_k$, 
completely negligible.   Corrections to the $\mu$ term are at most
of the order of $\mu b_0 a_0/M_k^2$, again, a negligible correction.
 
The radiative corrections within the supersymmetric theory are 
strongly restricted by the constraints of supersymmetry.  All component
fields within supermultiplet receive the same radiative corrections. By
the non-renormalization theorem~\cite{GRS}, the
superpotential receives no corrections.  
The result of this theorem constrains only 
the leading term in a Taylor series in external momenta, but, since these
diagrams are evaluated at external momenta of order $\MSUSY$, terms that
depend on external momenta 
are suppressed by powers of $\MSUSY/M_k$ and can 
be ignored.  Then the most general 
effective
Lagrangian obtained by integrating out the $N_k$ multiplets will
have the form
\beq
 \L_\eff =   \int d^4\theta \, \bar L_i (1+\delta Z_L)^{ij} L_j  
          +\int d^4\theta\,  \bar E_i (1+\delta 
Z_{E})^{ij} 
E_j+ \int d^2\theta \, W + H.c.
\eeq{effsusyL}
Since the Lagrangian is real-valued,  the matrices $(\delta Z_L)^{ij}$
and $(\delta Z_E)^{ij}$ must be Hermitian to all 
orders in perturbation theory.
 Note that while $ (\delta Z_L)^{ij}$ receives 
off-diagonal 
corrections at the one-loop level, $ (\delta Z_E)^{ij}$
receives off-diagonal elements only at the  two-loop level 
because $E$ does not have any flavor number violating coupling.

To generate a lepton electric dipole moment, we require a flavor-diagonal
matrix element of an electromagnetic form factor to have an 
imaginary part~\cite{offdiagonal}.
However, the radiative corrections from the supersymmetric Lagrangian,
 treated to first order,
will be proportional to the matrices $\delta Z_L$ and  $\delta Z_E$.
Since the diagonal elements of a 
Hermitian matrix are real, none of these  corrections,
acting alone, can induce a lepton electric dipole moment.  This 
is an important constraint, which we will continue to follow through our
analysis.

The soft supersymmetry breaking part of the  
Lagrangian receives corrections proportional to the supersymmetry breaking 
parameters.  However, the form is still quite constrained.
The most general effective Lagrangian has the form
\beqa
 \L_{SSB } &=& - (m_0^2 + \delta m_{\tilde{L}}^{2})_{ij} 
\s{L}^\dagger_i \s{L}_j
- (m_0^2 + \delta m_{\tilde{E}}^{2})_{ij}
\s{E}^\dagger_i \s{E}_j \CR
& & -  (a_0 Y_{\ell i} \delta_{ij} + \delta {\cal A}^{ij})
          \epsilon_{\alpha\beta} H_{1\alpha} \tilde{E}_i 
\tilde{L}_{j\beta} + H.c.
\eeqa{effsoftsusyL}
Since $\L_{SSB }$ is Hermitian,  $(\delta 
m^{2}_{\tilde{E}})_{ij}$ and ($\delta
m^{2}_{\tilde{L}})_{ij}$ must be  Hermitian matrices 
to all orders in perturbation theory.
The $A$ term can in general receive non-Hermitian contribution.  However,
we will show in Appendix A that, up to order $Y_\nu^4$, the corrections 
to the $A$ term have the form
\beq
    \delta{\cal  A}^{ij} = a_0 Y_{\ell i} \delta Z_A^{ij} \ ,
\eeq{formofA}
where $\delta Z_A$ is Hermitian.  Here again, the form of the radiative
corrections as Hermitian matrices limits their ability to contribute
to electric dipole moments.

To work with the effective Lagrangian written in
\leqn{effsusyL} and \leqn{effsoftsusyL}, it is useful to
bring the lepton and slepton fields into a canonical
normalization by rescaling by $(1+\delta Z)^{-1/2}$.  
Then the superpotential becomes \beq W = - [(1+ \delta
Z_E)^{-1/2}]^{ki}Y_{\ell i} [(1+ \delta Z_L)^{-1/2}]^{ij}
             \epsilon_{\alpha\beta} H_{1\alpha}
                     E_k L_j
\eeq{newsuper}
and the soft terms become 
\beqa
 \L_{SSB \eff} &=& - [(1+ \delta Z_L)^{-1/2}(m_0^2 + 
\delta m^{2}_{\tilde{L}})
  (1 + \delta Z_L)^{-1/2}]^{ij} \s{L}^\dagger_i \s{L}_j  
\CR
& &  - [(1+ \delta Z_E)^{-1/2}(m_0^2 +
\delta m^{2}_{\tilde{E}})
  (1 + \delta Z_E)^{-1/2}]^{ij} \s{E}^\dagger_i \s{E}_j
\cr 
& & \hskip -0.5in
 -  a_0  [(1 + \delta Z_E)^{-1/2}Y_\ell (1 + \delta 
Z_A)(1 + \delta Z_L)^{-1/2}]^{ij} 
                  \epsilon_{\alpha\beta} 
          H_{1\alpha} \tilde{E}_i \tilde{L}_{j\beta} + H.c.
\eeqa{neweffsoftsusyLone}

One more step is needed.  To identify the mass basis for leptons, we
need to  re-diagonalize the lepton Yukawa coupling.  Decompose the 
coefficient of \leqn{newsuper} into a product of a unitary matrix, a 
real positive diagonal matrix, and another unitary matrix:
\beq
      [(1 + \delta Z_E)^{-1/2}Y_{\ell } (1 + \delta 
Z_L)^{-1/2}]_{ij} = 
[(1 + \delta V)^T {\cal Y}_\ell
               (1 + \delta U)]_{ij} 
\eeq{Ydecomp}
Then $(1+\delta V)^T$ can be absorbed into the 
superfields 
$E$ and 
$(1+\delta U)$ can be absorbed into the superfields $L$.  The soft 
supersymmetry breaking terms now take a form similar to 
\leqn{effsoftsusyL}:
\beqa
 \L_{SSB } &=& - (m_0^2 + \Delta m_{\tilde{L}}^{2})_{ij} 
\s{L}^\dagger_i \s{L}_j
- (m_0^2 + \Delta m_{\tilde{E}}^{2})_{ij}
\s{E}^\dagger_i \s{E}_j \CR
& & -  a_0 {\cal Y}_{li} (\delta_{ij} + \Delta Z_A^{ij})
          \epsilon_{\alpha\beta} H_{1\alpha} \tilde{E}_i 
\tilde{L}_{i\beta} + H.c.
\eeqa{effsoftsusyLtwo}
where
\beqa
  (m_0^2 + \Delta m_{\tilde{L}}^{2}) &=&  [(1+\delta U)
(1+ \delta Z_L)^{-1/2}(m_0^2 + \delta m_{\tilde{L}}^{2})
  (1 + \delta Z_L)^{-1/2}(1+\delta U)^{-1}] \CR
(m_0^2 + \Delta m_{\tilde{E}}^{2}) & =& [(1+\delta V)
(1+ \delta Z_E)^{-1/2}(m_0^2 + \delta m_{\tilde{E}}^{2})
  (1 + \delta Z_E)^{-1/2}(1+\delta V)^{-1}]\CR
a_0 {\cal Y}( 1 + \Delta Z_A) &=& a_0 {\cal Y}_{\ell} 
     (1 + \delta U)(1 + \delta Z_L)^{1/2}
                       (1 + \delta Z_A)
(1 + \delta Z_L)^{-1/2} (1 + \delta U)^{-1}\ . \CR
\eeqa{newsoftcoeffs}

At this point, the only signs of CP-violation from the neutrino Yukawa
couplings occur in the coefficient functions listed in \leqn{newsoftcoeffs}.
It is still true that the first two
 coefficient functions are Hermitian matrices
with real diagonal elements, and that the diagonal elements of the 
$A$ term coefficient are real through two-loop order (order $Y_\nu^4$).  
For the mass matrices, this result is obvious. For the $A$ term an additional
slightly technical argument is needed, which we give in Appendix~B.

This implies that, through order $Y_\nu^4$, we cannot obtain a contribution 
to the lepton electric dipole moments from any individual term in 
\leqn{effsoftsusyLtwo}.  However, we can obtain a matrix with an imaginary
part by taking the product of two different matrices from 
\leqn{effsoftsusyLtwo}.  For example, 
\beq
    C_i =      \Im \left[     \Delta Z_A  \Delta m_{\tilde{L}}^2 \right]_{ii}
\eeq{tryprod}
can have nonzero diagonal elements.
 Since both 
matrices are Hermitian, 
this quantity can be written more illustratively as 
\beq
            C_i  =  {1\over 2i}  \left(  [ \  \Delta Z_A\ ,\  
\Delta m_{\tilde{L}}^2 \  ]
                     \right)_{ii}   \ . 
\eeq{Cidef}
Note that to compute 
$C_i$ to order of $Y_\nu^4$, it suffices to calculate
$ \Delta Z_A$ and $\Delta m_{\tilde{L}}^2$ to the
one-loop level.
Through  two-loop order, this is the only structure in the theory that 
can contribute to a lepton electric dipole moment.

At three-loop order, products of $\Delta m_{\tilde{E}}^2$ 
with the other matrices
in \leqn{effsoftsusyLtwo} can give additional contributions of a new 
structure.  A specific CP-violating quantity that will be important to us
is 
\beq
            D_i  =  
\Im  \left(  (\Delta m_{\tilde{E}}^2)^T\, 
m_\ell\  \Delta m_{\tilde{L}}^2   \right)_{ii}
\eeq{Didef}
This quantity also has a commutator structure, as we will see in Section 7.
It is smaller than \leqn{Cidef} by a factor of $Y_\ell^2/4\pi$.
Nevertheless, as we will see in Section 7, this term can give the dominant
contribution to lepton electric dipole moments in models with large
$\tan\beta$.   To obtain the contribution from this structure
of order $Y_\nu^4 Y_\ell^2$, it suffices to calculate  $\Delta m_{\tilde{E}}^2$
to two-loop order and  $\Delta m_{\tilde{L}}^2$ to one-loop order.

\begin{figure}
\begin{center}
\epsfig{file=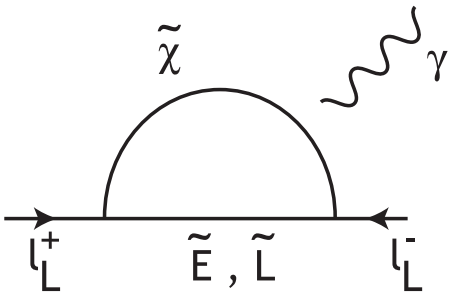, height = 1.5in}
\caption{The general form of the diagrams contributing to the 
EDM of a charged lepton $\ell$.  The photon line should be attached 
at all possible positions in the diagram.}
\label{fig:EDM}
\end{center} \end{figure}

\begin{figure}
\begin{center}
\epsfig{file=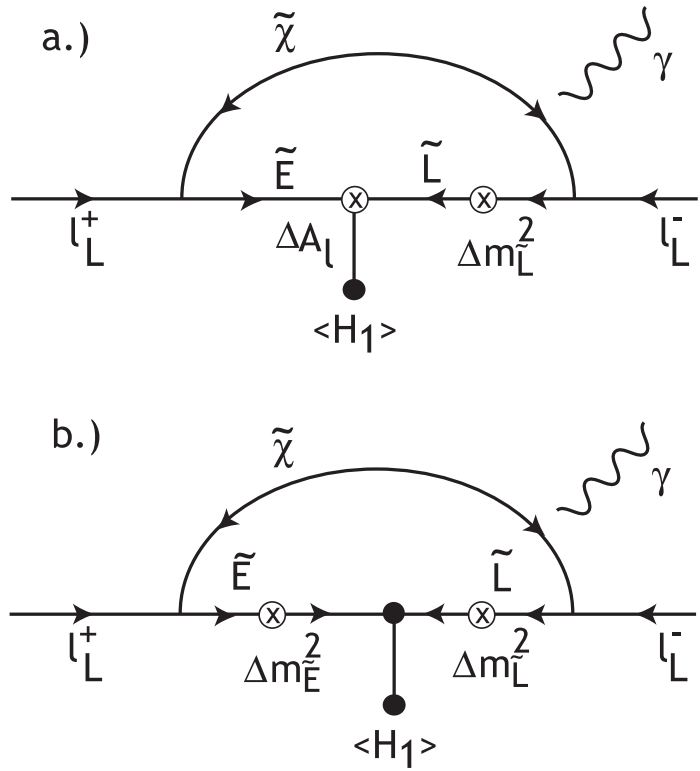, height = 4in}
\caption{Diagrams giving the 
dominant contribution to EDM of 
charged lepton $\ell$  (a) for small $\tan 
\beta$, (b) for large $\tan\beta$.}
\label{fig:EDM1}
\end{center} \end{figure}

We can be somewhat more concrete about how the structures $C_i$ and $D_i$
arise from Feynman diagrams.  Contributions to the lepton EDM's come
from diagrams of the general form of Fig.~\ref{fig:EDM}, 
in which a right-handed
lepton and is converted to a left-handed lepton through a photon vertex
diagram.  A lepton line runs through the diagram, and the matrices
\leqn{newsoftcoeffs} appear as insertions on this line.  By the arguments
just given, we need to consider contributions with two separate insertions.
The product \leqn{Cidef} 
comes uniquely from diagrams of the form of
Fig.~\ref{fig:EDM1}(a), with the photon inserted in all possible positions
on the lepton line.  The product \leqn{Didef} comes from diagrams of the
form of Fig.~\ref{fig:EDM1}(b).   In the latter diagram, the left-right 
mixing contributes the factor of $m_\ell$.  We will evaluate these diagrams
in Sections 6 and 7, respectively.


\section{One-loop corrections}

\begin{figure}
\begin{center}
\epsfig{file=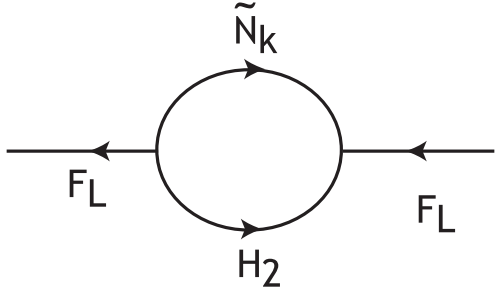, height = 1.5in}
\caption{Diagram giving the field strength renormalization of the
supermultiplet $L_i$. In this and the next few figures, we treat $F$
components as independent fields; the $F$ terms of $N_k$ multiplets
have the propagators \leqn{theprops}.}
\label{fig:deltas}
\end{center} \end{figure}

To estimate the lepton electric dipole moments at order $Y_\nu^4$, we 
should next compute
$\Delta Z_A$ and $\Delta m_{\tilde{L}}^2$.  According to the arguments of the 
previous section, only the leading-order contributions are needed. To 
this order
\beq
          \Delta m_{\tilde{L}}^2 =  \delta m_{\tilde{L}}^2 
- m_0^2 \delta Z_L \qquad
    \Delta Z_A = \delta Z_A
\eeq{computedeltas}
The factor $\delta Z_L$ is most easily computed as the one-loop correction
to the $F_L$ field strength.  There is only one 
diagram, 
shown in Fig.~\ref{fig:deltas}; its value is
\beq
   (\delta Z_L)^{ij}  =  (Y_\nu^{ki})^* Y_\nu^{kj} 
\int {d^4 p_E \over (2\pi)^4}
       {1\over          p_E^2 (p_E^2 + M_k^2)} \ ,
\eeq{deltaZLval}
where $p_E$ is a Euclidean momentum after Wick rotation.

\begin{figure}
\begin{center}
\epsfig{file=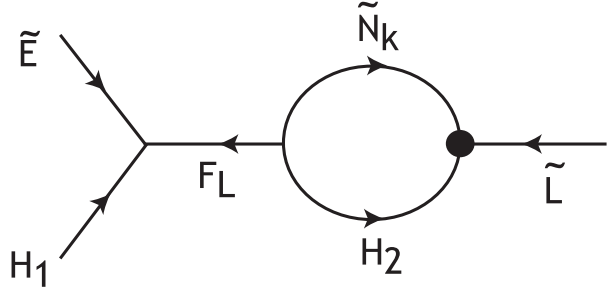, height = 1.5in}
\caption{Diagram giving the one-loop radiative correction to the 
vertex $A_\ell$. The heavy dot is an $A_\nu$ vertex.} 
\label{fig:deltaa}
\end{center} \end{figure}

The factor $\delta Z_A$ arises from the diagram shown in
 Fig.~\ref{fig:deltaa}. The vertex marked with a heavy dot 
is an $A_\nu$ vertex.  
The value of the diagram is
\beq
  a_0 Y_{\ell i} (\delta Z_A)^{ij}  =  - a_0 Y_{\ell i}
           (Y_\nu^{ki})^* Y_\nu^{kj} \int {d^4 p_E 
\over (2\pi)^4}
            {1\over     p_E^2 (p_E^2 + M_k^2)} \ .
\eeq{deltaZAval}
The tensor structure
 is exactly the same as in \leqn{deltaZLval}. This 
fact is used in Appendix~B.

\begin{figure}
\begin{center}
\epsfig{file=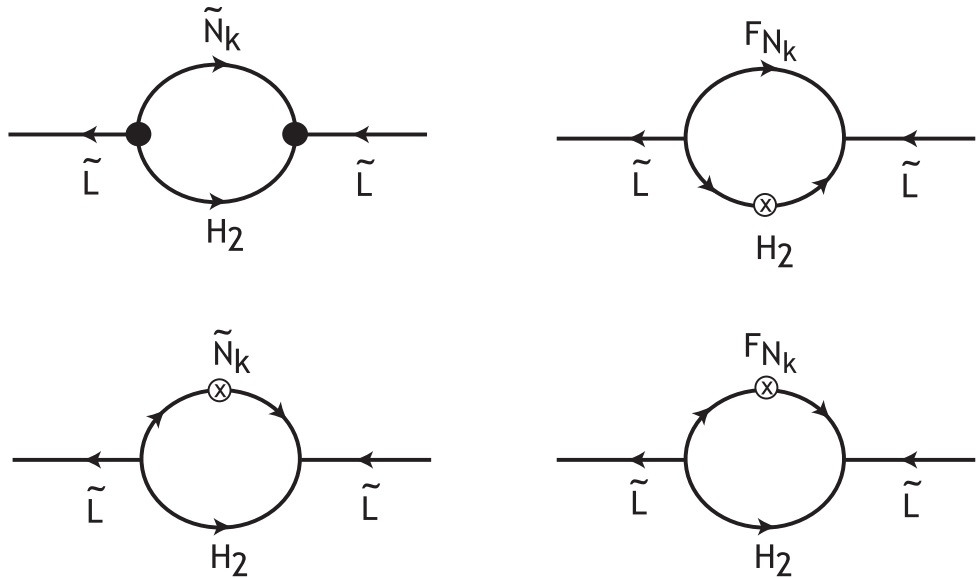, height = 3in}
\caption{Diagrams giving the one-loop corrections to the supersymmetry
breaking $\tilde L$ mass term. The heavy dot is an $A_\nu$ vertex; the
marked insertion is a soft mass term $m_0^2$.}
\label{fig:deltam}
\end{center} \end{figure}

The matrix $\delta m^2_L$ arises from the four diagrams 
shown in 
 Fig.~\ref{fig:deltam}. The first diagram has two $A_\nu$ vertices; the
other three have supersymmetry breaking 
mass insertions.  It should be noted that there is a contribution in which
$m_0^2$ in inserted into the $F_N$ propagator, which results from the 
mixing of $F_N$ with $\s N$ through the Majorana mass term.
The final result is
\beq
   (\delta m_{\tilde{L}}^2)^{ij}  =  -  (Y_\nu^{ki})^* Y_\nu^{kj} 
\int {d^4 p_E 
\over (2\pi)^4} 
\left[ {m_0^2 + a_0^2\over p_E^2 (p_E^2 + M_k^2)}  + 
      {m_0^2\over (p_E^2 + M_k^2)^2 }-{m_0^2 
M_k^2 \over p_E^2(p_E^2 + M_k^2)^2 }\right] \  ,
\eeq{deltaZmval}
which is quite similar to \leqn{deltaZLval} and \leqn{deltaZAval}, except
that some terms appear with two massive propagators.  The small difference
in structure between \leqn{deltaZmval} and the earlier equations will be
significant.

As we have explained in Section 2, we regularize these diagrams by 
dimensional regularization and minimal subtraction at the scale 
$\MGUT$.  This gives
\beqa
 \delta Z_L^{ij}  &=&   {1\over (4 \pi)^2}(Y_\nu^{ki})^* 
Y_\nu^{kj} 
              \left[\log {\MGUT^2\over M_k^2 
}+1\right] \CR 
 \delta Z_A^{ij}  &=&   - {1\over (4 \pi)^2}(Y_\nu^{ki})^* 
Y_\nu^{kj} 
             \left[ \log {\MGUT^2\over M_k^2 } 
+1\right]\CR 
 (\delta m_{\tilde{L}}^2)^{ij}  &=&  - {2m_0^2\over (4 
\pi)^2}(Y_\nu^{ki})^* Y_\nu^{kj} \left[
               \log {\MGUT^2\over M_k^2 }\right] 
-{a_0^2\over (4 \pi)^2}(Y_\nu^{ki})^* Y_\nu^{kj}
             \left[\log {\MGUT^2\over M_k^2 }+1\right] 
\ .
\eeqa{allZs}
so that
\beq
\Delta m_{\tilde{L}}^2{}^{ij} =  - {1\over (4\pi)^2}
(Y_\nu^{ki})^* Y_\nu^{kj} 
   \left(  m_0^2 [ 3\log {\MGUT^2\over M_k^2 }+1] 
+a_0^2[\log {\MGUT^2\over M_k^2 }+1]
\right) \ . 
\eeq{finalDmL}
and $\Delta Z_A = \delta Z_A$.

Now some significant simplifications appear.  
First,  in evaluating
\leqn{Cidef}, we can drop any terms in $\Delta m_{\tilde{L}}^2$ that are
proportional to the tensor structure of $\delta Z_A$.  Thus, we can 
replace
\beq
\Delta m_{\tilde{L}}^2{}^{ij} \to  - {1\over (4\pi)^2}
(Y_\nu^{ki})^* Y_\nu^{kj}(-2  m_0^2) \ .
\eeq{substitute}
  Second, after making this 
simplification, we can drop any terms in $\delta Z_A^{ij}$ that are 
proportional to the structure $(Y_\nu^{ki})^* Y_\nu^{kj}$.  In particular,
we can change $\MGUT$ inside the logarithm to any other value that is 
independent of $k$.
  We then find
\beq
C_i  ={m_0^2\over (4\pi)^4} {(\, [ {\bf Y}_0, {\bf
Y}_1]\,)_{ii}\over i}  \ ,\eeq{commutators} 
where 
\beq
  ({\bf Y}_0)^{ij} = (Y_\nu^{ki})^* Y_\nu^{kj}  \qquad 
   ({\bf Y}_1)^{ij} = (Y_\nu^{ki})^* Y_\nu^{kj} \log{M^2_N\over M_k^2} \ .
\eeq{Yforms}
As is explained just above, the expression for $C_i$ 
actually does not depend on 
the parameter $M_N$. It is 
convenient to choose $M_N$ to be the geometric mean of the $M_k$ to 
minimize the individual logarithms that appear in \leqn{Yforms}.

Our final result for $C_i$ is simple and cutoff-independent.  However,
we remind the reader that this result is derived in  the simple picture
in which we ignore threshold effects at the GUT scale and regulate
diagrams using the $\bar{DR}$ scheme.  Because of the major cancellations
that occurred in the simplification of $C_i$,  these threshold
corrections, which depend in a model-dependent way on 
 GUT-scale physics, can be of the
same order of magnitude as \leqn{commutators}.

\section{Comparison to the RGE approach}

It is remarkable that, to order $Y_\nu^4$, the only contribution to 
the lepton EDM comes from the invariant $C_i$ and that there is no
contribution from $\Im[A_\ell]$.  This conflicts with previous results on
lepton EDMs given by EHRS~\cite{ellisteam} and Masina~\cite{isabella}.  
In this 
section, we will compute the leading logarithmic contributions
to  $\Im[A_\ell]$ using the renormalization group method and demonstrate
explicitly that they cancel.  At the end of the section, we will compare
our analysis to that of \cite{ellisteam} and \cite{isabella}.

To carry out the renormalization group analysis, we must integrate
the RGEs from an initial condition at 
$\MGUT$ to the heaviest
$N$ mass, $M_3$, then from $M_3$ to $M_2$, then 
from $M_2$ to $M_1$. This procedure is 
valid only if $M_1\ll M_2 \ll M_3$. Let us define \beq
          t(Q)  = {1\over (4\pi)^2} \log Q \ .
\eeq{tdef}
and 
\beq
    t_{3} = t(\MGUT) - t(M_3)\ , \qquad    
t_{2} = t(M_3) - t(M_2)\ , \qquad
             t_{1} = t(M_2) - t(M_1)     \ . 
\eeq{morets}
For a hierarchical spectrum of masses, 
 we expect this procedure to reproduce the 
results of two-loop 
calculations up to the order
$Y^4_\nu  t^2$. 

It is very important to write the RGEs in such a way that the right-handed
neutrino thresholds are accounted correctly.  There are two aspects to this.
First, one should,  at each stage of 
integration,   
project out those $N_i$'s that have masses 
above the scale at which
the RGE is being evaluated.  To discuss this, it is useful to 
introduce projectors  $P_3 = 1$, $P_2=$ diag(1, 1, 0), $P_1=$ diag(1, 0, 0), 
projecting onto the $N$ mass eigenstates that are still active 
as we integrate through the various thresholds.  Second, one should be careful
to keep the matrices $Y_\ell$ and $M$ diagonal,
 at least when heavy particles are integrated out.
 
\begin{figure}
\begin{center}
\epsfig{file=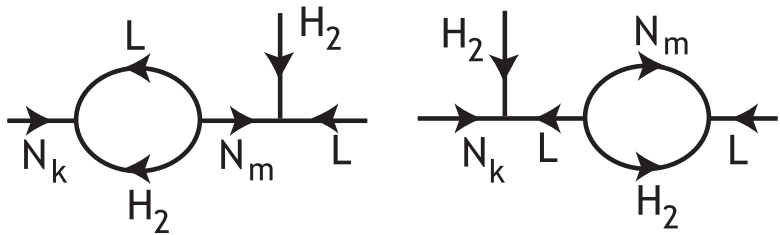, height = 1.2in}
\caption{Diagrams giving the terms proportional to $Y_\nu^2$ 
in the  RGE evolution of the neutrino 
Yukawa coupling.}
\label{fig:myRGEex}
\end{center} \end{figure}

We found it surprising that it is necessary to worry about off-diagonal 
terms in $M$, and 
so we would like to illustrate this with an example.
In the appendix of \cite{isabella}, the RGE for the 
neutrino Yukawa coupling is given as 
\beq
   {d Y_\nu\over dt} = 3 Y_\nu Y_\nu^\dagger P_a Y_\nu  +  \cdots
\eeq{firstRGE}
The contribution on the right-hand side arises from the 
diagrams shown in Fig.~\ref{fig:myRGEex}.  The projector eliminates 
contributions from the right-handed neutrinos with mass $M_k > Q$.
Consider, in particular,
integrating this equation down to a $Q$ such that $M_2 < Q < M_3$. Let
$t_Q = t(M_3) - t(Q)$.  Then the integration gives
\beq
Y_\nu(Q)= Y_\nu(\MGUT)+ 3 Y_\nu Y_\nu^\dagger Y_\nu t_3
           + 3  Y_\nu Y_\nu^\dagger P_2 Y_\nu t_Q \ .
\eeq{firstintegral}
However, direct calculation of the diagrams in Fig.~\ref{fig:myRGEex} with
Euclidean external momenta with $|Q^2| \ll M_3^2$ gives
\beq
Y_\nu(Q)= Y_\nu(\MGUT)+ 2 Y_\nu Y_\nu^\dagger P_2 Y_\nu (t_3 + t_Q)
              +  Y_\nu Y_\nu^\dagger Y_\nu t_3 
                +  Y_\nu Y_\nu^\dagger Y_\nu t_Q \ ,
\eeq{realintegral}
since in the first diagram the contribution from $N_3$ in the internal
line labelled $N_m$ has a propagator proportional to 
$1/(Q^2 + M^2)$ and so  is suppressed for  $Q^2 \ll M_3^2$.

The direct calculation is correct.  The problem is that, at this level,
the application of the renormalization group method is incomplete.
 The neutrino mass
matrix also acquires off-diagonal terms from the RGE.  For $Q \gg M_3$ 
\beq
   {d M\over dt } =  2 (Y_\nu Y_\nu^\dagger) M + 2 M (Y_\nu Y_\nu^\dagger)^T
              + \cdots \ .
\eeq{massRGE}
Thus, we should, first, integrate all of the RGEs down to a scale of 
the order of $M_3$, second, diagonalize the mass matrix $M$ at this scale
and rewrite the couplings in this new basis, 
third, integrate out its largest eigenvalue, and, finally, use the
rotated couplings as the initial conditions for the stage of integration
from $M_3$ to $M_2$.  

It is not difficult to see that this prescription
precisely 
eliminates the term that does not appear
in \leqn{realintegral} from \leqn{firstintegral}.  
The first diagram in Fig.~\ref{fig:myRGEex}
modifies the neutrino Yukawa coupling by a field strength renormalization
factor $Y_\nu \to Z_N^{-1/2} 
Y_\nu$.
The equation \leqn{massRGE} induces a similar modification in the mass
matrix,  $M \to  Z_N^{-1/2} M  (Z_N^{-1/2})^T$.  The $Z_N$ factors are the 
same in the two
expressions due to the nonrenormalization theorem.
When we now diagonalize $M$ at the scale $M_3$, the change of basis 
cancels the off-diagonal 1--3 and 2--3 elements of $Z_N^{-1/2}$ that 
affect $Y_\nu$.

To control this effect, one must
integrate through all three thresholds by carefully
solving the RGE for all couplings and mass terms.  However, here we only
wish to develop expressions for the effective couplings to order 
$t^2$, in order to check the results of the previous sections.  For this,
 it is 
easier and more direct to use the following procedure:  First, we 
integrate the 
renormalization group equations for the couplings. Then we 
identify terms that 
correspond to diagrams such as the first one in Fig.~\ref{fig:myRGEex} with
decoupling internal lines, and we remove these contributions by hand.

We should be careful also to remove diagrams with  
intermediate $F_N$ lines, since $F_N$ also
decouples, as we see from the last line of \leqn{theprops}.  
One-loop
diagrams involving the supersymmetry breaking $a_0$ term can 
produce  mixing of $L$ and $F_L$  or $N$ and $F_N$, for example, as in the
diagram shown in Fig.~\ref{fig:deltaa}.  In the RGE evolution of $A_\nu$,
we encounter a term in which the intermediate
line is $F_{N3}$. In this contribution,
 the off-diagonal terms decouple and should be removed
at the same time that we remove intermediate $N_3$ lines.
  Unless this is done, one cannot see the 
complete cancellation of $\Im[A_{\ell i i}]$ that we will present below.
In an RGE analysis, this step requires treating $F_N$ as a separate field
in the Lagrangian and  diagonalizing the 
$N^* F_N$ quadratic terms generated through RGE evolution.

With this insight into how to treat self-energy terms in the
RGEs, we can integrate the RGEs for coupling constants through the three
thresholds.  The renormalization group 
equations for coupling constants are as given the Appendix of~\cite{isabella},
\beqa
 {d Y_\nu\over dt }  &=&  3 Y_\nu K_a + \cdots \CR
      {d Y_\ell\over dt }  &=&  Y_\ell K_a +  
\cdots \CR
      {d A_\nu\over dt }  &=&  4 \tilde{K_a} A_\nu+5A_\nu K_a   + \cdots 
\CR
      {d A_\ell\over dt }  &=&  2 Y_\ell (Y_\nu)^\dagger A_\nu +
               A_\ell K_a +  \cdots \CR
{d m_{\tilde{L}}^2 \over dt }  &=& 
\{m_{\tilde{L}}^2\ , K_a \}+2(Y_\nu^\dagger P_a m_{\tilde N}^2 P_a Y_\nu 
+m_{H_u}^2K_a+A_\nu^\dagger P_aA_\nu)+ \cdots \CR
{d m_{\tilde{E}}^2 \over dt }  &=&   2(m_{\tilde{E}}^2Y_\ell^\dagger 
Y_\ell+Y_\ell^\dagger 
Y_\ell m_{\tilde{E}}^2)+4(Y_\ell^\dagger 
m_{\tilde{L}}^2Y_\ell+m_{H_u}^2Y_\ell^\dagger Y_\ell +A^\dagger_\ell 
A_\ell) + \cdots  \ , 
\eeqa{RGEsofIsabella}
where $K_a = Y_\nu ^\dagger P_a Y_\nu$ and  
$\tilde K_a = Y_\nu Y_\nu ^\dagger P_a$ and the subscript $a$ specifies 
the energy scale. Note that $  m_{\tilde N}^2$ is a supersymmetry 
breaking mass  and should not be mistaken for a
 large supersymmetric neutrino mass 
$M_i$. The terms not written explicitly in \leqn{RGEsofIsabella} are 
terms with flavor structures such as 
$Y_\nu \cdot \mbox{tr}[Y_\nu Y^\dagger_\nu]$ that do not contribute to 
EDMs.  Using the prescription that we have explained above,
we find different results from those found previously.  When we solve 
for $A_\ell$ and for $Y_\ell$
at a scale 
much smaller than $M_1$, we find that the imaginary parts of  $Y_\ell$
and $A_\ell/a_0$ are identical and are equal to 
\beq
   \Im \left[  Y_\ell (K_3 K_2 t_3 t_2 + K_3 K_1 
t_3 t_1 + K_2 K_1 t_2 t_1)
          \right]\ .
\eeq{finalIM}

Now one more step is needed.  As in \leqn{Ydecomp}, we need to choose a new
basis for the leptons in which $Y_\ell$ is real diagonal after taking into 
account the radiative corrections due to the $N_i$. 
Since the imaginary parts of  $Y_\ell$
and $A_\ell/a_0$ found at the previous stage are 
identical, this completely removes the 
imaginary part of $A_\ell$, in agreement with our analysis in Section 3.

It is not clear to us how the analyses of \cite{ellisteam} and 
\cite{isabella} found nonzero diagonal terms of $\Im[A_\ell]$ at this
order.  The discussion in \cite{ellisteam} does not discuss the 
issue of re-diagonalizing $Y_\ell$ and $M$.
On the other hand, \cite{isabella}
writes the renormalization group equations in a way that explicitly
takes  into account
the decoupling of heavy states, as  we have noted above.
In addition, the calculations done in this paper keep 
$Y_\ell$ and $M$ diagonal by adding terms to these RGEs following the
`rotating basis' prescription of Brax and Savoy~\cite{Brax}.  The full
RGEs considered are not written explicitly in \cite{isabella}, but 
nevertheless they are  used to generate
the results that are quoted there~\cite{isabellanote}.
One possible difficulty is that this analysis might not remove the  terms
with decoupling $\VEV{N^* F_N}$ propagators that we have discussed above 
\leqn{RGEsofIsabella}. 

The observation that lepton EDMs are proportional to the commutator of
${\bf Y}_0$ and ${\bf Y}_1$ is the most important result of the analysis 
of EHRS~\cite{ellisteam}.  Once this result has been found, it is 
straightforward
to obtain the correct order of magnitude for the contribution to lepton 
EDMs from the phases of neutrino Yukawa couplings.  Thus, the qualitative
dependence of EDMs on the underlying supersymmetry parameters is given
correctly in this paper, even though the actual terms that produce the
lepton EDMs are different.
   
\section{Electric dipole moments}

We are now ready to obtain the actual expression for the lepton EDMs by 
evaluating the class of diagrams shown in 
Fig.~\ref{fig:EDM}.

A general diagram of the form of Fig.~\ref{fig:EDM} evaluates to the form
\beq
      -ie v^T(p') c \sigma^{\mu\nu} q_\nu u(p) \cdot \left[- 
       {1\over m_{\ell i}} (F_{2i} + i F_{25i} ) \right] \ , 
\eeq{vertexdef}
where $F_2$ is the usual magnetic moment form factor and $i$ indexes the
lepton flavor.  The lepton EDM  is then given by 
\beq
       \vec d_i =  - e  F_{25i}  {\vec S\over m_i} = 
(1.9\times 10^{-11} \,
        \mbox{e\ cm})\cdot  F_{25i} \cdot {m_e \over m_{\ell i }} \cdot
  \hat S \ .
\eeq{edmval}
where $\vec S$ is the spin of the lepton and $\hat S = \vec S/(\hbar/2)$.

We would like to find a contribution to the EDM proportional to $C_i$
in \leqn{Cidef}.  For this, we should find a vertex diagram that depends on
both $A_\ell$ and $m_{\tilde{L}}^2$ and insert the flavor-violating 
corrections 
found in Section 3.  The only such diagram is shown in 
Fig.~\ref{fig:EDM1}(a).
The value of this diagram, as a contribution to 
$F_2+iF_{25}$, is
\beqa
   F_2+iF_{25} &=&  {\alpha\over 2 \pi}\sum_a
  \left({V_{01a} \over c_w} \right)
\left({V_{01a} \over c_w} + {V_{02a}\over 
 s_w }\right) 
(A_\ell - \mu\tan\beta)  m_{\ell i}^2 m_a \CR 
& &  \cdot  \int^1_0 \, dz \, \int^1_0 \, dx \,
{z (1-z)^2\over ( z m_a^2 + (1-z) (x m_{\tilde{E}}^2 
+ (1-x) m_{\tilde{L}}^2))^2} \ .
\eeqa{Ftwo}
In this expression, $V_0$ is the unitary
 matrix that diagonalizes the 
neutralino mass matrix:
\beq
        \s{b}^0  =  \sum_a V_{01a} \s{\chi}^0_a  \qquad   
\s{w}^0  =  \sum_a V_{02a} \s{\chi}_a^0  \ ,
\eeq{Vdef}
$m_a$ are the neutralino mass eigenvalues (with signs),
$c_w = \cos\theta_w$,
$s_w = \sin\theta_w$.  

The renormalization-group running of the soft 
supersymmetry breaking masses from the GUT scale to the 
electroweak scale corrections gives large but flavor-in\-de\-pen\-dent
corrections to the $\tilde{E}$ and $\tilde{L}$ masses proportional to the
GUT-scale gaugino masses.   These terms do not contribute to the 
flavor-violating effects that give the dipole matrix element an imaginary 
part, but they should be taken into account in the denominator of 
\leqn{Ftwo} in evaluating this imaginary part.  Thus, we have written
\leqn{Ftwo} as depending on the electroweak-scale values of these masses
$m_{\tilde{E}}$ and $m_{\tilde{L}}$.  The full expression \leqn{Ftwo}
can be checked against many 
papers on lepton dipole 
moments, for example, \cite{LA,LB,LC}.

Starting from \leqn{Ftwo}, we replace $A_\ell$ by $a_0 Y_\ell \delta Z_A$, 
and we include one mass insertion in the $\s{L}$ line by acting on 
the integral with 
\beq
               \Delta m_{\tilde{L}}^2 {\del\over \del m_{\tilde{L}}^2}
\eeq{insertM}
A schematic version of this analysis for general flavor-violating
perturbations is described, for example, in 
\cite{LFMS}.
In our model, we take the indicated derivative of \leqn{Ftwo}, 
assemble the structure $(\Delta Z_A \Delta m_{\tilde{L}}^2)$, and 
replace the imaginary part of this object by 
$(i C_i)$ as given in \leqn{Cidef}. 
We thus obtain an expression for
the lepton electric dipole moment of the form of \leqn{edmval}, where
\beq
   F_{25i} =  {2\alpha\over (4 \pi)^5 }\sum_a
  \left({V_{01a} \over c_w} \right)
\left({V_{01a} \over c_w} + {V_{02a}\over 
 s_w }\right)  {m_0^2 m_{\ell i}^2 a_0 m_a \over |m_a|^6}\ 
              { ([\, {\bf Y}_0 \, , \, {\bf Y}_1 \, 
])_{ii}\   \over i}
g({m_{\tilde{L}}^2\over m_a^2}, {m_{\tilde{E}}^2\over m_a^2})  \ , 
\eeq{Ftwovalue}
where $g(x_L, x_E)$ is given in Appendix C.  For comparison with the 
results of the next section, we might write this result alternatively as
\beqa
   F_{25i} &=&  {2\alpha\over (4 \pi)^5 }\sum_a
  \left({V_{01a} \over c_w} \right)
\left({V_{01a} \over c_w} + {V_{02a}\over 
 s_w }\right)  {m_{\ell i}^2 a_0 m_a \over |m_a|^6}
g({m_{\tilde{L}}^2\over m_a^2}, {m_{\tilde{E}}^2\over m_a^2})\CR
& & \hskip 0.5in \cdot 
\Im[ (Y_\nu^{ki})^* Y_\nu^{kj} (Y_\nu^{mj})^* Y_\nu^{mi}] 
 \cdot (-2m_0^2  \log{M_N^2\over M_k^2}) \ .
\eeqa{secondFtwo}

There is a curious consequence of this result that follows from the
fact that the trace of any commutator is zero.  If this effect is the 
only source of the lepton EDM, we expect that
\beq
d_e/m_e+d_\mu/m_\mu +d_\tau/m_\tau=0 \ .
\eeq{sumrule}
It is unclear to us how this simple formula could be tested to the required
accuracy.

\section{Electric dipole moments for large $\tan \beta$}

Masina \cite{isabella} 
has argued that, for large $\tan\beta$, a different contribution
to the lepton EDM can be the dominant one.  
Looking back at the diagrams
of Fig.~\ref{fig:EDM} and \ref{fig:EDM1} studied in the 
previous section, we see
that it is advantageous at large $\tan\beta$ 
to drop $A_\ell$ and keep
instead the term $\mu\tan\beta$.  We still need a second loop correction
to combine with $\Delta m_{\tilde{L}}^2$, but this can come from inserting 
 $\Delta m_{\tilde{E}}^2$ in the right-handed slepton propagator.  Since
 $\Delta m_{\tilde{E}}^2$ arises at order $Y_\ell^2 Y_\nu^2$, 
the new 
diagram has a size relative to the previous one of 
\beq
      {Y_\ell^2 \over (4\pi)^2} \tan\beta  \approx 
 { m_\tau^2 \over 8 \pi^2 v^2}{\tan^3 \beta \over \sin^2 \beta} \ , 
\eeq{newlargebetaval}
where $v = 246$ GeV, assuming that the $\tau$ lepton dominates the 
intermediate states in the matrix product.  We will see in a moment that,
whereas all large logarithms of $\MGUT$ cancelled out of the expression for 
lepton EDM in Section~6, the contribution of the large $\tan \beta$
region is enhanced by two powers of this large logarithm.
 As a result, the terms we will compute in this section 
 can dominate over those we discussed in Section~6 for 
$\tan\beta > 10$. 

\begin{figure}
\begin{center}
\epsfig{file=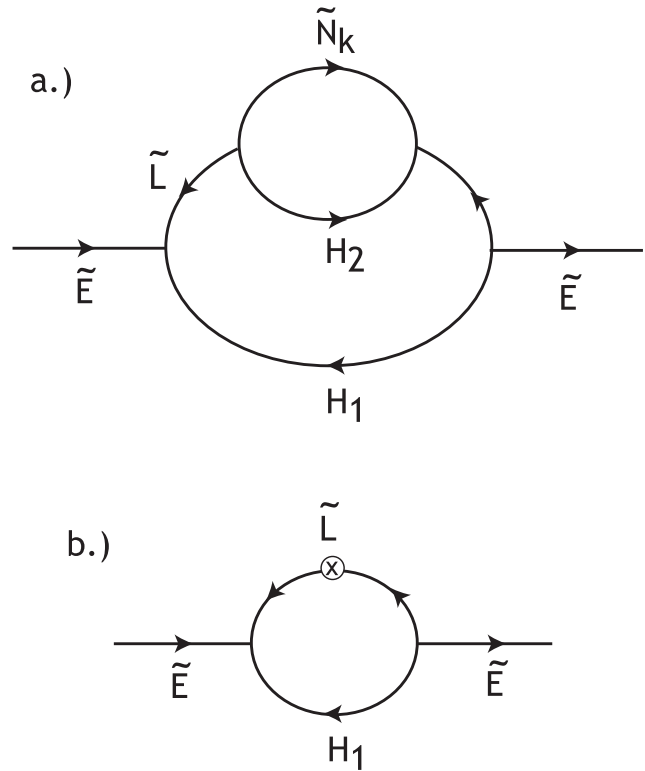, height = 4in}
\caption{The structure of the two-loop contributions to 
$\Delta m_{\tilde{E}}^2$ that involve the neutrino Yukawa couplings.
All possible particles from each supermultiplet should be put on each 
line of each diagrams:  (a) proper two-loop contributions; (b) one-loop
diagrams containing one-loop counterterms for $\Delta m_{\tilde{L}}^2$.}
\label{fig:twoloopform}
\end{center} \end{figure}

To evaluate this contribution, we need to work out the mass insertion
$\Delta m_{\tilde{E}}^2$ in \leqn{effsoftsusyLtwo}.   To begin, we must 
compute
$\delta Z_E$ and $\delta m_{\tilde{E}}^2$ up to the two-loop level.  
We need only compute those two-loop
diagrams that contain the maximum number
of $Y_\nu$ vertices, since only these diagrams will contain factors of
the CP violating phases needed for a contribution to the EDMs.

Consider first $\delta Z_E$.   The one- and two-loop diagrams 
contributing to the field strength renormalization give
\beqa 
\delta Z_E^{ji} & = & 2 Y_{\ell i}^2 \delta_{ij} \int {d^4 
p_E\over 
        (2\pi)^4} {1\over 
          (p_E^2)^2 }   \CR   &  &   
 -2 Y_{\ell i} (Y_\nu^{ki})^* Y_\nu^{kj} Y_{\ell j} \int {d^4 
p_E\over (2\pi)^4}
  \int {d^4 k_E\over (2\pi)^4}{1\over 
          (p_E^2) [(k_E-p_E)^2 + M_k^2] (k_E^2)^2 }  \ .
\eeqa{deltaZEtotwo}
The two-loop contribution of order $Y_\ell^2 Y_\nu^2 $ comes from 
a diagram of the topology of Fig.~\ref{fig:twoloopform}(a).
Notice that the indices of $\delta Z_E$ are transposed.  This is 
appropriate because, in the figure, 
the direction of the arrows is reversed on the $E$ lines.
In addition to the two-loop diagram,
 there is a one-loop diagram involving the one-loop
$\delta Z_L$ counterterm.  This diagram has topology shown in 
Fig~\ref{fig:twoloopform}(b) and has the value
\beq
\delta Z_E^{ji} = 
 2 Y_{\ell i} (Y_\nu^{ki})^* Y_\nu^{kj} Y_{\ell j} \int {d^d 
k_E\over (2\pi)^d} {1\over (k_E^2)^2 } {1\over (4\pi)^2} {1\over \epsilon}\ ,
\eeq{couterterm}
where $\epsilon = (4-d)/2$.

The contributions to $\delta m_{\tilde{E}}^2$ from one-loop diagrams
and from two-loop diagrams of the form of Fig.~~\ref{fig:twoloopform}(a)
are given by
\beqa
\delta m_{\tilde{E}}^2{}^{ji} & = & - 2 Y^2_{\ell i} 
\delta_{ij} 
\int {d^d p_E\over 
        (2\pi)^d} {2m_0^2 + a_0^2\over 
          (p_E^2)^2 }   \CR   &  &  + 
 2\, Y_{\ell i} (Y_\nu^{ki})^* Y_\nu^{kj} Y_{\ell j} \int {d^d 
p_E\over (2\pi)^d}
  \int {d^d k_E\over (2\pi)^d} 
{1 \over p_E^2[(k_E-p_E)^2 + M_k^2]^2(k_E^2)^2} \CR
& & \hskip 1.0in  \cdot
\left(5 m_0^2 + 4 a_0^2 
 - {2 m_0^2 M_k^2 \over 
          (k_E-p_E)^2 +M_k^2 }\right) \CR & & 
 -2 Y_{\ell i} (Y_\nu^{ki})^* Y_\nu^{kj} Y_{\ell j} \int {d^d 
k_E\over (2\pi)^d} {1\over (k_E^2)^2 } {1\over (4\pi)^2} {1\over \epsilon}
(5m_0^2 + 4a_0^2)\ ,
\eeqa{ingredients}
Again, we only consider corrections involving the $Y_\nu$ that
will contribute to the EDMs.
The first line of \leqn{ingredients} gives the complete one-loop contribution. 
The
second line gives the two-loop contribution proportional to $Y_\ell^2 
Y_\nu^2$.
This contribution comes from diagrams of the topology of 
Fig.~\ref{fig:twoloopform}(a). 
  To compute $\delta m_{\tilde{E}}^2$, we put $\s E$ on the 
external lines and
insert $m_0^2$ into the propagators or $a_0^2$ into the vertices in all
possible ways.  The final piece comes from diagrams of the topology of
Fig.~\ref{fig:twoloopform}(b) with the counterterms associated with the 
one-loop corrections to $Z_L$, $m_{\tilde{L}}^2$, and $Z_A$.

Note the order of the 
indices
on  $\delta m_{\tilde{E}}^2{}^{ji}$ in  these 
contributions;
this reflects the reversed direction of arrows 
on the external lines in  
Fig.~\ref{fig:twoloopform}.  Also, note that the integrals over $k_E$
contain superpartners $\tilde L$, $\tilde H_1$ that have masses of the TeV
scale rather than the right-handed neutrino scale.  These integrals are 
potentially infrared divergent, and we will replace $(k_E^2) \to 
(k_E^2 + \MSUSY^2)$ to regulate this divergence.

To compute the final mass insertion $\Delta m_{\tilde{E}}^2$, we must 
now make the 
redefinitions in \leqn{newsoftcoeffs}.  If we expand in the Yukawa 
couplings, 
we find
\beq
  \Delta m_{\tilde{E}}^2 = (\delta m_{\tilde{E}}^2  - m_0^2 \delta Z_E)  + 
       [ \, \, \delta V\ ,  
(\delta m_{\tilde{E}}^2 - m_0^2 \delta Z_E) \, ]
  \ + \cdots\ ,
\eeq{finalDelta}
where $\delta V$ is introduced in (\ref{Ydecomp}).
To give a nonzero diagonal element in \leqn{Didef}, we must expand the 
quantities in the first term to order $Y_\ell^2 Y_\nu^2$.  In the second 
term,
we will obtain a nonzero contribution to \leqn{Didef} by taking
 the one-loop expressions for $\delta m_{\tilde{E}}^2$ and 
$\delta Z_E$ together
with the one-loop expression for $\delta V$ that follows from the $\delta 
Z_L$
contribution to \leqn{Ydecomp}.  Note, while the flavor-independent 
gauge corrections to $\delta m_{\tilde{E}}^2$ and  $\delta Z_E$
commute with $\delta V$, the first terms in Eqs. 
\leqn{deltaZEtotwo} and \leqn{ingredients}, although 
flavor-conserving, do not commute with $\delta V$. That is why 
we have dropped the gauge correction in Eqs.  
\leqn{deltaZEtotwo} and \leqn{ingredients} while we have kept the 
$Y_\ell^2$ terms.
 According to the 
above equation,
\beq
  \delta V  Y_\ell^2 - Y_\ell^2 \delta V = Y_\ell 
(\delta Z_L)^* Y_\ell \ .
\eeq{VYrelation}
If we recognize that the one-loop 
expressions for $\delta m_{\tilde{E}}^2$ and $\delta Z_E$ are proportional 
to $Y_\ell^2$,
we can use this expression to evaluate the second term of \leqn{finalDelta}.
Inserting the value of $\delta Z_L$ given in \leqn{deltaZLval} and
transposing the matrix,  we find
a contribution of the same structure $Y_\ell Y_\nu^\dagger Y_\nu Y_\ell$ that
we have in the other contributions to $ (\Delta m_{\tilde{E}}^2)^T$.

Our final result for $\Delta m_{\tilde{E}}^2$ is then
\beqa
\Delta m_{\tilde{E}}^2{}^{ji} & =&   2 Y_{\ell i} (Y_\nu^{ki})^* Y_\nu^{kj}
  Y_{\ell j} \CR
  & & \hskip-0.5in
 \cdot \biggl\{   \int {d^d k_E\over (2\pi)^d}  {d^d p_E\over (2\pi)^d} 
{1 \over (k_E^2 + \MSUSY^2)^2 p_E^2 ((k_E-p_E)^2 + M_k^2)} 
\biggl( 6 m_0^2 + 4 a_0^2 -  {2 m_0^2 M_k^2\over (k_E-p_E)^2 + M_k^2} \biggr)
\CR
  & & \hskip 0.2in -   \int {d^4 k_E\over (2\pi)^4} {1\over 
 (k_E^2 + \MSUSY^2)^2 } \cdot {1\over (4\pi)^2}\bigl( {1\over \epsilon}\bigr)
     (6 m_0^2 + 4 a_0^2) \CR
  & & \hskip 0.2in - \bigl( {1\over (4\pi)^2} \log{\MGUT^2\over \MSUSY^2} 
\bigr) \bigl( {1\over (4\pi)^2} (\log{\MGUT^2\over M_k^2}+1 ) \bigr) 
       (3m_0^2 + a_0^2) \biggr\} \ .
\eeqa{bigdeltamE}

The two-loop integrals are standard forms that are evaluated, for example,
in the Appendices of \cite{vBV} and \cite{SMevals}.  
Using these results, we 
find for the off-diagonal elements of $\Delta m_{\tilde{E}}^2$
\beqa
\Delta m_{\tilde{E}}^2{}^{ji} &=&   {2\over (4\pi)^4}
        Y_{\ell i} (Y_\nu^{ki})^* Y_\nu^{kj} Y_{\ell j} \CR
   & &\cdot  \biggl\{  (6m_0^2 + 4a_0^2) \biggl[ 
               \half\log^2 {\MGUT^2 \over M_k^2}
     + \log{\MGUT^2 \over M_k^2} \log{M_K^2 \over \MSUSY^2} \CR
  & & \hskip 1.0in + \log {\MGUT^2 \over \MSUSY^2} + \half - {\pi^2\over 6}
          \biggr] \CR
 & & \hskip 0.1in  - 2m_0^2 \log {M_k^2\over \MSUSY^2}  - 
    (3m_0^2 + a_0^2) \log {\MGUT^2\over \MSUSY^2} ( \log {\MGUT^2\over M_k^2}
     +1 ) \biggr\} \ . 
\eeqa{mne}
This formula is the exact result to order $Y_\ell^2 Y_\nu^2$ with ultraviolet
regularization by minimal subtraction at $\MGUT$.  It does not
assume that the right-handed neutrino masses are hierarchial.  The 
dependence
on $\MSUSY$, with terms of at most one logarithm,
 is consistent with renormalization group evolution from the 
heavy neutrino scale to the weak scale.  The leading logarithmic terms
in this expression are in precise agreement with the result of 
Masina~\cite{isabella}. 

The dominant contribution to the EDMs for large $\tan\beta$ is now found
by inserting both $(\Delta m_{\tilde{E}}^2)^T$ and 
$\Delta m_{\tilde{L}}^2$
 into the vertex diagram
as shown in Fig.~\ref{fig:EDM1}(b). The imaginary part of the diagram is
proportional to the structure \leqn{Didef}.  The contributions to this 
formula have up to three powers of logarithms.  Just as in the 
evaluation of $C_i$, we can take advantage of the fact that we are computing
the imaginary part of the product of Hermitian matrices, which picks out the
antisymmetric product of these matrices.  In this case, the result contains
the structure
\beq
\Im[ (Y_\nu^{ki})^* Y_\nu^{kj} m_{\ell j}^2(Y_\nu^{mj})^* Y_\nu^{mi}]   \ ,
\eeq{Yfourstructure}
contracted by a function of $M_k$ and $M_m$.
Note that the structure in (\ref{Yfourstructure}) is 
antisymmetric in 
the right-handed neutrino flavor indices $k$ and $m$.
When we antisymmetrize the expression contracted with 
this structure, 
the leading term with $\log^3 (\MGUT^2/M_k^2)$ 
cancels out. However, while for $C_i$ the next subleading logarithm also 
cancels out, here it does not and so, unlike the previous case,
 the final 
result will depend on $\MGUT$.
More precisely, we find
\beqa
D_i &=&  \Im \Biggl\{  {4\over (4\pi)^6 }{m_{\ell 
i}\over v^2 \cos^2\beta} 
 (Y_\nu^{ki})^* Y_\nu^{kj} m_{\ell j}^2(Y_\nu^{mj})^* Y_\nu^{mi} \CR
& & \cdot \biggl[ m_0^4 
\Bigl(  9 \log{\MGUT^2\over M_N^2} \log {\MGUT^2\over M_k^2}
        \log {M_N^2\over M_k^2} 
        + 9 \log^2 {M_N^2\over M_k^2} \log {M_N^2\over M_m^2}\CR
 & & \hskip 0.6in 
  + 6  \log{\MGUT^2\over M_N^2} \log {M_N^2\over M_k^2} 
       + 3 \log^2  {M_N^2\over M_k^2}
            + (7-\pi^2) \log {M_N^2\over M_k^2}  \Bigr) \CR
& & \hskip 0.1in + a_0^2 m_0^2 
\Bigl(  9 \log{\MGUT^2\over M_N^2} \log {\MGUT^2\over M_k^2}
        \log {M_N^2\over M_k^2} 
        + 9 \log^2 {M_N^2\over M_k^2} \log {M_N^2\over M_m^2}\CR
 & & \hskip -0.1in 
    + 14  \log{\MGUT^2\over M_N^2} \log {M_N^2\over M_k^2} 
       + 5 \log^2  {M_N^2\over M_k^2}
    + 4  \log{M_N^2\over M_k^2} \log {M_N^2\over \MSUSY^2} 
            + (7-3\pi^2) \log {M_N^2\over M_k^2}  \Bigr) \CR
& & \hskip 0.1in + a_0^4 
\Bigl(  2 \log{\MGUT^2\over M_N^2} \log {\MGUT^2\over M_k^2}
        \log {M_N^2\over M_k^2} 
        + 2 \log^2 {M_N^2\over M_k^2} \log {M_N^2\over M_m^2}\CR
 & & \hskip 0.6in 
    + 4  \log{\MGUT^2\over M_N^2} \log {M_N^2\over M_k^2} 
       + 2 \log^2  {M_N^2\over M_k^2}
            + (2- {2\over 3}\pi^2) \log {M_N^2\over M_k^2}  \Bigr) \biggr]
 \Biggr\} \ . \CR
\eeqa{Dicompute}
The parameter $M_N$ is a mean right-handed neutrino mass.  The precise
definition of this mass is unimportant, because, as in 
\leqn{commutators},  the various factors of 
$M_N$ cancel out of \leqn{Dicompute} when we use the antisymmetry of 
the structure $\Im[Y_\nu^4]$.   It is convenient to choose $M_N$ as the 
geometric mean of the $M_k$ to minimize the individual logarithms in 
\leqn{Dicompute}.

If the right-handed neutrino masses are strongly hierarchial,
as was assumed by \cite{isabella}, \leqn{Dicompute} is enhanced by three
large logarithmic factors.  The formula we have given here is valid
for any right-handed neutrino spectrum; for a spectrum without 
large hierarchies, the leading term still has two large logarithms.
 We also confirm 
the result of \cite{isabella}
that the logarithmic dependence
on $\MSUSY$ cancels in the leading order of logarithms, though a small
dependence does remain in a subleading term.  The coefficient of our 
leading term is identical to that found by Masina.

\begin{figure}
\begin{center}
\epsfig{file=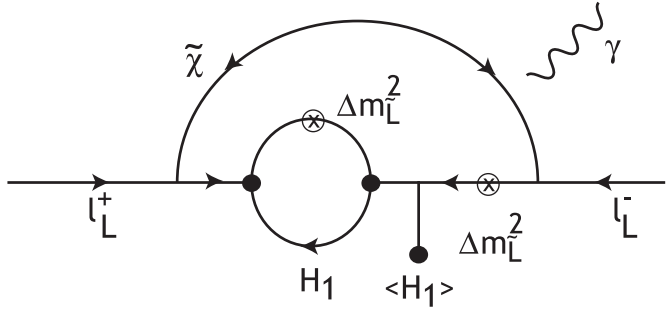, height = 1.6in}
\caption{Sub-dominant diagram contributing to EDMs. 
The vertices marked by heavy dots are $A_\ell$ vertices,
and the marked insertion is a one-loop correction to 
$m_{\tilde{L}}^2$.  All of the momenta flowing through the
indicated loops are of order
$\MSUSY$.}\label{fig:aaaa}
 \end{center} 
\end{figure}

From this expression we obtain lepton EDMs of the form of \leqn{edmval}
with 
\beqa
   F_{25i} &=&  - {8\alpha\over (4 \pi)^7 }
  \left({V_{01a} \over c_w} \right)
\left({V_{01a} \over c_w} + {V_{02a}\over 
 s_w }\right)\,  {\mu m_{\ell i}^2 m_a \over |m_a|^8 v^2}\
 {\tan\beta\over \cos^2\beta} \
\CR
& & \hskip 0.6in \cdot 
\Im[ (Y_\nu^{ki})^* Y_\nu^{kj} m_{\ell j}^2 (Y_\nu^{mj})^* Y_\nu^{mi}] 
 \  h({m_{\tilde{L}}^2\over m_a^2}, {m_{\tilde{E}}^2\over m_a^2}) \CR
& & \hskip -0.3in \cdot \biggl[ ( 9 m_0^4  + 9 a_0^2 m_0^2 + 2 a_0^4) 
       \Bigl(\log{\MGUT^2\over M_N^2} \log {\MGUT^2\over M_k^2}
        \log {M_N^2\over M_k^2}  + 
           \log^2{M_N^2\over M_k^2} \log {M_N^2\over M_m^2} \Bigr)
\biggr] \ . \CR
\eeqa{finalFtwo}
where $h(x_L,x_E)$ is given in Appendix~C.  
In the above formula,  we
have kept only the leading logarithmic terms, that is, terms with 
 two large logarithms in the case of a general right-handed neutrino
mass spectrum and with three large logarithms in the case of a 
hierarchial mass spectrum.  If we wish to give an
expression valid, in the general case,
 at the level of one large logarithm, we should include 
the corrections to $D_i$ from the TeV threshold, replacing the $\MSUSY$ by the
actual masses of $\tilde L$ and $H_1$.  At the same time, we must include
an additional contribution, shown in
Fig.~\ref{fig:aaaa}, involving a two-loop integral with momenta at the 
TeV scale. An analysis to this accuracy is beyond the scope of 
this paper.

\section{Discussion}

In this paper, we have re-evaluated the contributions from neutrino
Yukawa couplings to the lepton EDMs.  In contrast to previous studies,
we have shown that, in the mass basis of  
charged leptons, up to two-loop level, neutrino Yukawa couplings 
do not induce an imaginary part to the diagonal elements of
the $A_\ell$ term. However,  complex
neutrino Yukawa couplings can create EDMs for charged leptons through 
differences in the renormalization of the $A_\ell$ terms and the 
slepton masses terms,
through the diagrams shown in Figs.~\ref{fig:EDM} and 
~\ref{fig:EDM1}.
Our expressions for the lepton EDMs have the same 
structure in terms of the
neutrino Yukawa couplings 
as those
previously given in \cite{ellisteam} and \cite{isabella}.  However, the
form of the integrals contributing to $F_{25i}$ is different because
there is an extra mass insertion.  Further, the overall size of the effect 
is decreased from the previous estimates, especially in the region of 
low $\tan\beta$.

There is an important test for the origin of lepton EDMs in the neutrino
sector.  While complex $a_0$ and $\mu$ induce EDMs both for the 
charged leptons and for the neutron, 
effects from the neutrino sector give zero EDM for the neutron
while making nonzero contributions for the leptons.  However, an EDM 
present only for leptons could in principle arise from an
 imaginary part to the $A_\ell$ coefficient or the 
neutrino $B$ term as well as from loop effects involving $Y_\nu$.  It is
interesting to compare the magnitudes of the effects from loop level or
tree level CP-violating contributions.

In the low $\tan \beta$ region, the effect that we have computed in 
\leqn{Ftwovalue} gives a lepton EDM of the order of magnitude
\beq
d_i\sim 10^{-29}\, Y_\nu^4\,  \log {M_3^2 \over M_1^2} 
\left( {200 \ \ {\rm 
GeV} \over \MSUSY} \right)^2  \left( {m_{\ell i} 
\over m_e} \right) \ \mbox{e\ cm}. \
\eeq{firstdeval}
The effect from the large $\tan\beta$ region has a double logarithmic
enhancement with respect to this value.  If we estimate
$\log^2(\MGUT^2/M_N^2) \sim 200$, the effect that
we have computed in \leqn{finalFtwo} gives a lepton EDM of the 
order of magnitude
\beq
d_i\sim 10^{-29}\, \left({\tan\beta\over 10}\right)^3 \,
Y_\nu^4\,  \log {M_3^2 \over M_1^2} 
\left( {200 \ \ {\rm 
GeV} \over \MSUSY} \right)^2  \left( {m_{\ell i} 
\over m_e} \right) \ \mbox{e\ cm}. \
\eeq{seconddeval}
These estimates can be compared to the current best limit on the
electron EDM, $d_e < 1.6\times 
10^{-27}$ e~cm~\cite{EDMnow}.  To achieve an EDM close 
to the current bound, we would need to have   $Y_\nu^4 \log (M_1/M_3) \sim 
100$.  However, the experimental limit on the rate of $\mu\to e \gamma$
places a limit on the $Y_\nu$ matrix 
elements~\cite{antonio},
\beq
        Y_\nu^{*ke} Y_\nu^{k\mu} \log{\MGUT^2\over M_k^2}  < 0.1 \tan\beta \ ,
\eeq{YYlimit}
so it seems unlikely to have  such large values of the $Y_\nu$.  On the
other hand, the effect of the neutrino $B$ term leads  to a potentially
much larger estimate for electron EDM,
\beq
d_i\sim 10^{-27} {{\rm Im} (B_\nu) \over \MSUSY} Y_\nu^2 \left( {200
 \ \ {\rm GeV} \over \MSUSY }\right)^2 {m_{\ell i} 
\over m_e}\ \mbox{e \ cm}\ .
\eeq{BtermEDM}
This could easily saturate the present bound.
Also, we expect  $d_\mu\sim m_\mu/m_e d_e$, 
so if $d_e$ is close to its present bound, $d_\mu$ should also be 
observable in future muon storage ring experiments~\cite{Aysto}.
If complex Yukawa couplings are the only source of CP-violation and $Y_\nu 
\sim 1$, the electron EDM should still be observed in the 
next generation of experiments, which aim for sensitivity to 
$d_e \sim 10^{-29}$ e~cm~\cite{futureEDM}.

Over the longer term, CP-violating effects of  complex neutrino Yukawa 
couplings can also 
be probed by lepton flavor oscillation in slepton production at 
colliders~\cite{nima},
and perhaps also in sneutrino-antisneutrino oscillation~\cite{GHaber}.
Better understanding of the systematics of leptogenesis can also 
play a role on constraining the neutrino Yukawa couplings.  All of this
information will complement the knowledge that we are gaining from
neutrino oscillation experiments to help us build a complete picture
of the neutrino flavor interactions.

\Acknowledgements

We are grateful to Yuval Grossman, Helen Quinn, M. M. Sheikh-Jabbari 
and Alexei Smirnov
 for very useful discussions of 
CP violation in supersymmetric theories.  We also thank Isabella
Masina for an instructive correspondence.

\appendix
\section{Corrections to the lepton $A$ term}

In this Appendix, we will show that the contributions to the 
parameter $\delta {\cal A}^{ij}$ in the effective Lagrangian
\leqn{effsoftsusyL} up to order $Y_\nu^4$
are of the form of \leqn{formofA} with 
$\delta Z_A$ a Hermitian matrix.

The structure \leqn{formofA} is easy to see. The A-term interaction
must have one factor of $Y_\ell$, with corrections coming from 
neutrino Yukawa couplings.  Among the three fields coupled by 
$Y_\ell$, only $L_j$ couples to $Y_\nu$.  So the corrections we are
must be 2-point diagrams attached to the $\s L_j$ external leg.

The difficult part is to argue that these diagrams add to a Hermitian
matrix.  For this point, we will give a formal argument and a diagrammatic
argument.  The formal argument is as follows:   We can generate the 
universal A-term in our model by multiplying the superpotential \leqn{theW}
by  a new chiral superfield $S$, which we will eventual set equal to 
$(1 + a_0 \theta^2)$.   With this 
structure, the superpotential is not renormalized.  However, a correction
to the A-term of the 
structure \leqn{formofA} can be generated by if loop corrections generate
new terms in the K\"ahler potential of the form
\beq
       \delta K^n =   \delta Z_{ij}^n  (\bar S S)^n \bar L_i L_j \ .
\eeq{newKahler}
Then $\delta Z_A$ is a linear combination of the  $\delta Z^n$.  
However, since
 \leqn{newKahler} is a correction to the K\"ahler potential, $\delta Z^n$
must be Hermitian~\cite{Hitoshi}.

As an alternative  to this argument, 
we provide an explicit diagrammatic analysis up to 
order $Y_\nu^4$.  In particular, for the 
problem at hand,
we are interested in polynomials in $Y_\nu$ that contribute to the 
$A_\ell$ term and have a nonzero imaginary part.

A given diagram with only one $N$ line could, in
principle,
contain structures
$Y_\nu \cdot Y_\nu^*$, $Y_\nu\cdot Y_\nu$, or $Y_\nu^*\cdot
Y_\nu^*$. The vertex $A_\ell$ conserves the number of $H_2$ 
(in fact 
$H_2$ has no $A_\ell$ coupling). However, the vertices $A_\nu$
and $Y_\nu$ change the number of $H_2$ by one unit.
  Since we ignore the masses of $L$ and $H_2$
in diagrams
involving $N$, the $L$ and $H$ numbers are conserved by
internal propagators.
Therefore, any radiative correction to $A_\ell$
  has equal
numbers of $Y_\nu$
and $Y_\nu^*$.

Consider a diagram contributing to $A_\ell$ with only 
one 
$N$ line and $(n+m)$
$Y_\ell$ vertices.
From the above result, we see that the most general 
polynomial that can 
appear in such diagrams 
is
\beq
                     (Y_\ell^j)^n\sum_k 
(Y_\nu^{kj})^* Y_\nu^{ki}  f(M_k) (Y_\ell^i)^m
\eeq{oneloopYY}
whose diagonal elements are purely real. Notice 
that in the case of
one-loop diagram shown in Fig.~\ref{fig:deltaa}, 
$n=1$, $m=0$ and the matrix $\delta Z_A$ [defined in \leqn{formofA}] is 
Hermitian.
\begin{figure}
\begin{center}
\epsfig{file=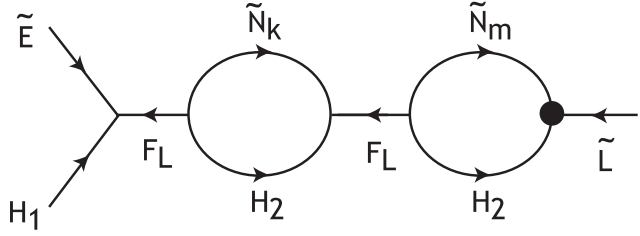, height = 1.2in}
\caption{A contribution to the renormalization of $A_\ell$ in 
two-loop order from two one-loop diagrams.}
\label{fig:1by1}
\end{center} \end{figure}


Now let us focus on two-loop diagrams with more than one 
$N$ line. In such diagrams four $A_\nu$-vertices 
are involved.
A contribution from a product of two one-loop diagrams, as shown
in Fig.~\ref{fig:1by1}, has the polynomial 
structure
\beq
    \sum_{mn} Y_\nu^{m\alpha} Y_\nu^{mk * }   Y_\nu^{nk} 
Y_\nu^{n\beta * } f_1(M_m)f_2(M_n)\ .
\eeq{twooneloopYY} 
It is non-trivial but easy to 
show that the functions $f_1$ and $f_2$ are of the 
same form. As a result, the matrix in 
\leqn{twooneloopYY} is 
Hermitian.

\begin{figure}
\begin{center}
\epsfig{file=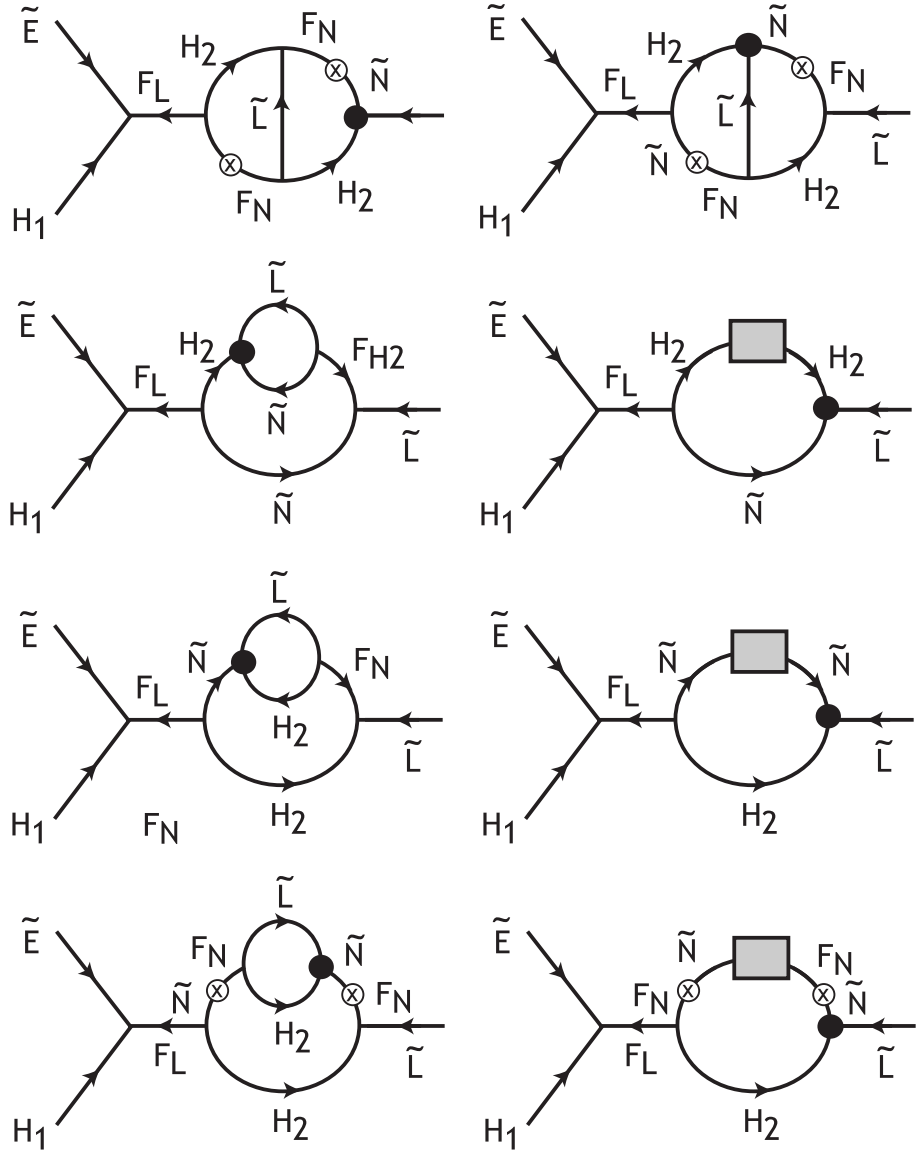, height = 6in}
\caption{Irreducible two-loop diagrams contributing
to $A_\ell$. The shaded boxes
represent the full one-loop propagator corrections from the $N$, $H_2$
supermultiplets.}
\label{fig:as} 
\end{center} \end{figure}

This brings us to irreducible two-loop diagrams 
contributing to $A$ terms.
The structures of these diagrams fall into three
categories: 1) they can be of the form 
\beq  \sum_{mn}  Y_\nu^{mi} Y_\nu^{mk * }   Y_\nu^{nk} 
Y_\nu^{nj * }g_1(M_n,M_m) \ ,  \eeq{nna}
2) they can be of the form
\beq  \sum_{mn} Y_\nu^{mi}(Y_\nu^{m j})^* 
Y_\nu^{nk}(Y_\nu^{nk})^*
g_2(M_n,M_m) \ ,  \eeq{nn}
3) or they can be of the form
\beq
 \sum_{m,n,k}   Y_\nu^{m i} Y_\nu^{mk}   (Y_\nu^{nk})^*
(Y_\nu^{nj})^*
g_3(M_m,M_n)  \ ,
\eeq{twoloopYY}
where $g_1$, $g_2$, $g_3$ are real functions of $M_n$ and $M_m$.
The structure shown in \leqn{nn} is 
manifestly Hermitian.
If the functions $g_1(M_m,M_n) $ and $g_3(M_m,M_n) $ are
symmetric
under $M_m \leftrightarrow M_n$, the structures appearing in \leqn{nna}
and \leqn{twoloopYY} will be Hermitian 	also.  It is not very obvious 
that these functions have the required symmetry.  But it is not 
difficult to show this by explicit examination of the diagrams.
All the relevant diagrams are shown in Fig.~\ref{fig:as}.
Since the momenta propagating in the loops are of 
order of $M_N$, we can neglect the external momenta, 
which for our purposes are of order of $\MSUSY$.
With this simplification, it can be seen
that all these diagrams
are symmetric under $M_m \leftrightarrow M_n$.
This completes the proof that, up to two-loop level, the 
diagonal elements of $A_\ell$ remain real. 

\section{Expansion of $\Delta {\cal A}$, eq. 
\leqn{newsoftcoeffs},  to order $Y_\nu^4$}

In Section 3, we claimed that the diagonal matrix element
\beq
\Bigl[ (1 + \delta U)(1 + \delta Z_L)^{1/2}
(1 + \delta Z_A)
(1 + \delta Z_L)^{-1/2} (1 + \delta U)^{-1} \Bigr]^{ii} 
\eeq{diagonmat}
is real through order $Y_\nu^4$.  To order $Y_\nu^2$, this is easy to see:
The matrix element is the matrix element of the sum 
\beq
  (\delta U + \half \delta Z_L +  \delta Z_A -   \half \delta Z_L - \delta U)
            =   \delta Z_A
\eeqn
and $\delta Z_A$ is Hermitian.  

Working to order $Y_\nu^4$, we first consider the separate contributions
of order $Y_\nu^4$ from each factor of 
$\delta Z_A$, $\delta Z_L$, and $\delta U$.  The factors of $\delta U$
cancel. The contributions from $\delta Z_A$ and  $\delta Z_L$  are
diagonal elements of Hermitian matrices and thus
manifestly real.

In addition, we must look at contributions in 
which two of these objects at a time are expanded to order $Y_\nu^2$.  To 
analyze these terms, we need the expressions for $\delta Z_L$ and $\delta Z_A$
given in \leqn{allZs}.  We also need an expression for $\delta U$.  The 
definition of  $(1+ \delta U)$ is that it diagonalizes the matrix
\beq
               (1 + \delta Z_L)^{-1/2} Y^2 (1 + \delta Z_L)^{-1/2} \ .
\eeqn
Using first-order quantum-mechanical perturbation theory, we find
that $(\delta U)_{ii} = 0$ and, for $i \neq j$,
\beq
            (\delta U)_{ij} =  {Y_i^2 + Y_j^2 \over Y_i^2 - 
Y_j^2 }
              {1\over 2(4 \pi)^2}(Y_\nu^{ki})^* Y_\nu^{kj}
              \left(\log {\MGUT^2\over M_k^2 }+1 \right) \ .
\eeqn
Then any diagonal element of a product of any two of  $\delta Z_L$,
 $\delta Z_A$, $\delta U$ is of the form of the quantity
\beq
      \left( (Y_\nu^{ki})^* Y_\nu^{kj} [
              \log {\Lambda^2\over M_k^2 }+1]\right)      
      \left( (Y_\nu^{pj})^* Y_\nu^{pi} 
              [\log {\Lambda^2\over M_p^2 }+1]\right) 
\eeqn
multiplied by a real-valued expression. No such term has an imaginary part.    

\section{Mass dependence of dipole matrix elements}

As we have explained in Sections 6 and 7, the dipole matrix elements that
contribute to lepton EDMs contain derivatives of the expression
\beqa
{1\over m_a^4} f(x_L,x_E) &=&   \int^1_0 \, dz \, \int^1_0 \, dx \,
{z (1-z)^2\over  (z m_a^2 + (1-z) (x m_{\tilde{E}}^2 
+ (1-x) m_{\tilde{L}}^2))^2}\CR
 &=&  \int^1_0 \, dz \, 
{z (1-z)\over m_{\tilde{E}}^2 - m_{\tilde{L}}^2}\left(
 {1\over  z m_a^2 + (1-z) m_{\tilde{L}}^2} -  {1\over  z m_a^2 
+ (1-z) m_{\tilde{E}}^2}\right) \ .
\eeqa{fintdef}
where $x_L = m_{\tilde{L}}^2/m_a^2$, $x_E = m_{\tilde{E}}^2/m_a^2$.  
This expression evaluates to
\beq
 f(x_L,x_E) =  {1\over 2} {1\over x_E- x_L}
      \left( {1 - x_L^2 + 2 x_L \log x_L \over (1-x_L)^3 } 
          - {1 - x_E^2 + 2 x_E \log x_E \over (1-x_E)^3 }  \right) \ . 
\eeq{fintvalue}

To make one insertion of $\Delta m_{\tilde{L}}^2$, we need
\beq
    {1\over m_a^6} g(x_L,x_E)   =  {\del\over \del m_{\tilde{L}}^2} 
{1\over m_a^4} f(x_L,x_E) \ . 
\eeq{gintdef}
This has the value
\beqa
   g(x_L,x_E) &=&  {1\over 2 (x_E-x_L)^2} \biggl(
  {1 - x_L^2 + 2 x_L \log x_L \over (1-x_L)^3 } 
          - {1 - x_E^2 + 2 x_E \log x_E \over (1-x_E)^3 } \biggr) \CR
 & & + {1\over 2(x_E- x_L)}  \biggl(
  {5 - 4 x_L - x_L^2 + 2(1+2x_L) \log x_L \over (1-x_L)^4 
}  \biggr) \ .
\eeqa{gintvalue}

To make one further insertion of $\Delta m_{\tilde{E}}^2$, we need
\beq
    {1\over m_a^8} h(x_L,x_E)   =  
{\del\over \del m_{\tilde{E}}^2} 
{1\over m_a^6} g(x_L,x_E) \ . 
\eeq{hintdef}
This has the value
\beqa
   h(x_L,x_E) &=&  - {1\over (x_E-x_L)^3} \biggl(
  {1 - x_L^2 + 2 x_L \log x_L \over (1-x_L)^3 } 
          - {1 - x_E^2 + 2 x_E \log x_E \over (1-x_E)^3 } \biggr) \CR
 & & - {1\over 2(x_E- x_L)^2}  \biggl(
  {5 - 4 x_L - x_L^2 + 2(1+2x_L) \log x_L \over (1-x_L)^4 
} \CR 
 & &   \hskip 0.9in
 + {5 - 4 x_E - x_E^2 + 2(1+2x_E) \log x_E \over (1-x_E)^4 
} \biggr) \ . 
\eeqa{hintvalue}

For $m_{\tilde{L}}^2$, $m_{\tilde{E}}^2 \gg m_a^2$, we find
\beq
  f(x_L,x_E) \approx {1\over 2 x_L x_E} \qquad 
  g(x_L,x_E) \approx - {1\over 2 x_L^2 x_E} \qquad 
  h(x_L,x_E) \approx {1\over 2 x_L^2 x_E^2}  \ ,
\eeq{finalvals}
as we might have expected.

\newpage

\end{document}